\documentclass{aa}
\usepackage[varg]{txfonts}
\usepackage{graphicx}
\usepackage{txfonts}
\usepackage[]{hyperref}
\usepackage{natbib}

\begin{document}

\title{A census of OBe stars in nearby metal-poor dwarf galaxies reveals a high fraction of extreme rotators}

\author{A. Schootemeijer\inst{1} \and D. J. Lennon\inst{2,3} \and M. Garcia\inst{4} \and N. Langer\inst{1,5} \and B. Hastings\inst{1,5} \and C. Sch\"{u}rmann\inst{1,5}
}

\institute{Argelander-Institut f\"{u}r Astronomie, Universit\"{a}t Bonn, Auf dem H\"{u}gel 71, 53121 Bonn, Germany\\  \email{aschoot@astro.uni-bonn.de} \and
Instituto de Astrofísica de Canarias, E-38200 La Laguna, Tenerife, Spain \and
Departamento de Astrof\'{i}sica, Universidad de La Laguna, E-38205 La Laguna, Tenerife, Spain \and Centro de Astrobiolog\'{i}a, CSIC-INTA. Crtra. de Torrej\'{o}n a Ajalvir km 4, E-28850 Torrej\'{o}n de Ardoz (Madrid), Spain
\and
Max-Planck-Institut für Radioastronomie, Auf dem Hügel 69, 53121 Bonn, Germany}



\abstract{The Early Universe, together with many nearby dwarf galaxies, is deficient in heavy elements.
The evolution of massive stars in such environments is thought to be affected by rotation. Extreme rotators amongst them tend to form decretion disks and manifest themselves as OBe stars. We use a combination of $UB$, GAIA, Spitzer, and Hubble Space Telescope photometry to identify the complete populations of massive OBe stars -- one hundred to thousands in number -- in five nearby dwarf galaxies. This allows us to derive the galaxy-wide fractions of main sequence stars that are OBe stars ($f_\mathrm{OBe}$), and how it depends on absolute magnitude, mass, and metallicity ($Z$).
We find $f_\mathrm{OBe} = 0.22$ in the Large Magellanic Cloud (0.5\,$Z_\odot$), increasing to $f_\mathrm{OBe} = 0.31$ in the Small Magellanic Cloud (0.2\,$Z_\odot$). In the so far unexplored metallicity regime below $0.2\,Z_\odot$, in Holmberg\,I, Holmberg\,II, and Sextans\,A, we also obtain high OBe star fractions of $ 0.27$, $ 0.27$, and $ 0.27$, respectively. These high OBe fractions, and the strong contribution in the stellar mass range which dominates the production of supernovae, shed new light on the formation channel of OBe stars, as well as on the preference of long-duration gamma-ray bursts and superluminous supernovae to occur in metal-poor galaxies.}


\maketitle

\section{Introduction} \label{sec:intro}

The early universe contained a small amount of heavy elements, and the same is true for dwarf galaxies \citep{Mateo98}. As such, nearby dwarf galaxies are uniquely suitable to study the early generations of massive stars --  the stars that could have reionized the universe \citep{Dayal18}, and whose feedback controlled early star formation.
As a result of the low abundance of heavy elements (referred to as metallicity, $Z$), massive stars (stars that are at least eight times more massive than the Sun) have weaker radiation-driven stellar winds \citep{Hainich15, Vink21}. Weak stellar winds lead to low angular momentum loss, and allow low-metallicity massive stars to maintain higher rotation velocities \citep{Mokiem07, Ramachandran19, Brott11} as well as higher masses prior to core collapse, favoring the formation of massive black holes.
Interestingly, superluminous supernovae \citep{Lunnan14}, long-duration gamma-ray bursts \citep{Savaglio09}, and ultra-luminous X-ray emission from compact sources \citep{Kaaret17} primarily occur in low-metallicity dwarf galaxies.
All of these have been linked to rapidly rotating massive stars \citep{Aguilera18, Marchant17}.

Rapid rotation is known to affect the structure, appearance, and evolution of massive stars in various ways. First, the closer stars are to their critical rotation velocity, the more they expand in the equatorial direction. Then, gravity darkening reduces their effective temperatures near the equator \citep{vonZeipel24, EspinosaLara11}, which makes rotating stars look redder. Second, as stars rotate at or close to their critical rotation velocity, they can acquire a decretion disk \citep{Rivinius13}. This disk emits more light in the red than in the blue, providing a second mechanism that makes an extremely rapidly rotating source intrinsically redder \citep{Martayan10}. Because of the emission lines that form in the disk, these stars are classified as ‘OBe’ stars \citep{Struve31}. Finally, rotation is thought to induce internal mixing. If efficient enough, this could lead to chemically homogeneous evolution, where stars remain compact and evolve to ever-higher temperatures and luminosities \citep{Maeder87, Langer92}.
This type of evolution would open an extra window to gravitational wave events caused by black hole  mergers \citep{deMink16, Marchant16}. Also, massive stars experiencing chemically homogeneous evolution could produce high-energy photons \citep{Szecsi15}, providing a potential explanation for nebular emission of, for example, doubly ionized helium that is observed in high-redshift, metal-poor galaxies \citep{Erb10}.

In the Milky Way and its satellite dwarf galaxies, the Large Magellanic Cloud \citep[LMC; 0.5\,$Z_\odot$][]{Trundle07} and the Small Magellanic Cloud \citep[SMC; 0.2\,$Z_\odot$][]{Korn00}, rotation velocities of large samples of massive stars have been inferred spectroscopically from line-broadening \citep[e.g.,][]{Mokiem06, Ramachandran19}. These studies found that the average rotation velocity of massive stars modestly increases with decreasing metallicity.
This trend might continue towards lower metallicity; but whether it does is uncertain, because of the currently scarce statistics in metal-poorer star-forming galaxies, 
resulting from how expensive spectroscopic observations of these distant systems are \citep{Telford21, Garcia21}.

In a method alternative to line broadening measurements for gauging stellar rotation, which is cheaper and does not suffer from inclination effects, one measures the fraction of OBe stars. There, we distinguish between two different OBe star fractions: the OBe star fraction measured in sub-samples of stars, and the galaxy-wide OBe star fraction.
A galaxy-wide OBe star fraction has so far never been measured.
In previous work, one approach to measure OBe star fractions was to use photometry to detect excess infrared (IR) emission from decretion disks of OBe stars in a sub-sample of stars with known spectral types in the Magellanic Clouds \citep{Bonanos09, Bonanos10}.
An alternative approach has been to use narrow-band H$\alpha$ photometry, which traces H$\alpha$ emission from the OBe star's disk, to measure the OBe star fraction in star clusters of the Magellanic Clouds \citep{Keller99, Wisniewski06, Iqbal13} and the Milky Way \citep{McSwain05}, as well as in parts of the Andromeda galaxy \citep{Peters20}. Measured cluster OBe star fractions tend to be higher at low metallicity, but within a galaxy the obtained results depend strongly on the cluster that is picked and which of its stars are included in the measurement (see below). We note that the studied clusters are often older than the maximum lifetime of massive stars -- which are the focus of this work -- such that they mainly probe longer-lived low-mass stars.
Also, measurements in clusters that are young enough to still contain massive stars might be biased because they can have lost OBe stars produced via the binary channel through supernova kicks \citep[as found by][]{Dallas22}.
Therefore, only a measurement of the galaxy-wide OBe star fraction can provide an unbiased approach. The goal of this work is twofold: i) measure this galaxy-wide fraction of bright OBe stars in the Magellanic Clouds, and ii) to extend the measurement of the galaxy-wide OBe star fraction to sub-SMC metallicity.

\section{Methods}


In this work, we make use of archival broad-band photometry. We use short-wavelength (near-UV/blue optical) data to identify hot stars -- which OBe stars are -- and longer-wavelength IR data to establish the presence of a disk. 
In this section, we will explain how we construct our data sets, of which we will show our analysis in the Results (Sect.\,\ref{sec:results}).
In Appendix\,\ref{sec:appb} we provide various tests, where we discuss in detail how complete our data sets are, and how well we can use them to identify OBe stars. 

\subsection{Data sets \label{sec:appa}}
\paragraph{Magellanic Clouds.}
For both the SMC and the LMC, with CDS-xmatch \citep[\url{http://cdsxmatch.u-strasbg.fr/};][]{Boch12, Pineau20} we cross-correlated sources from $UBVI$ catalogs \citep{Zaritsky02, Zaritsky04} to Spitzer-SAGE \citep{Meixner06} sources within 1". The resulting data files were then cross-correlated with GAIA EDR3 \citep{Gaia21} sources, with again a 1" cross-correlation radius. It can happen that a source from a $UBVI$ catalog is cross-matched to more than one source that would appear in our CMDs (i.e., brighter than $M_\mathrm{Gbp} = -2$, and a listed magnitude for each of the following filters: $U$, $B$, $Gbp$, $Grp$, $J$, and $[3.6]$). In those cases, we take the closest source. As a result, we discard 110 sources in the SMC and 353 in the LMC.
These cases are relatively few in number, since the SMC and LMC CMDs contain of the order of tens of thousands of sources. Therefore, we do not expect cross-correlation to the wrong sources to affect our results.

The cross-correlated catalogs are cleaned of foreground sources. For the SMC, the method is similar to the one from \cite{Schootemeijer21}: sources with a GAIA {\tt parallax\_over\_parallax\_error} value larger than five are removed, as well as sources that do not fit the proper motion criteria. More specifically, for the SMC we remove sources more than $2\sigma$ away from the SMC's bulk proper motion ($\mu$), for which we adopt the values from \cite{Yang19}: $\mu_\mathrm{ra} = 0.695 \pm 0.240$ mas/yr (right ascension) and $\mu_\mathrm{dec} = -1.206 \pm 0.140$ mas/yr (declination). For the LMC we take the same approach but for its bulk motion we adopt $\mu_\mathrm{ra} = 1.750 \pm 0.375$ mas/yr and $\mu_\mathrm{dec} = 0.250 \pm 0.375$ mas/yr, based on fig.\,16 from \cite{Gaia18b}.

\begin{figure*}[ht]
\centering
\includegraphics[width=\linewidth]{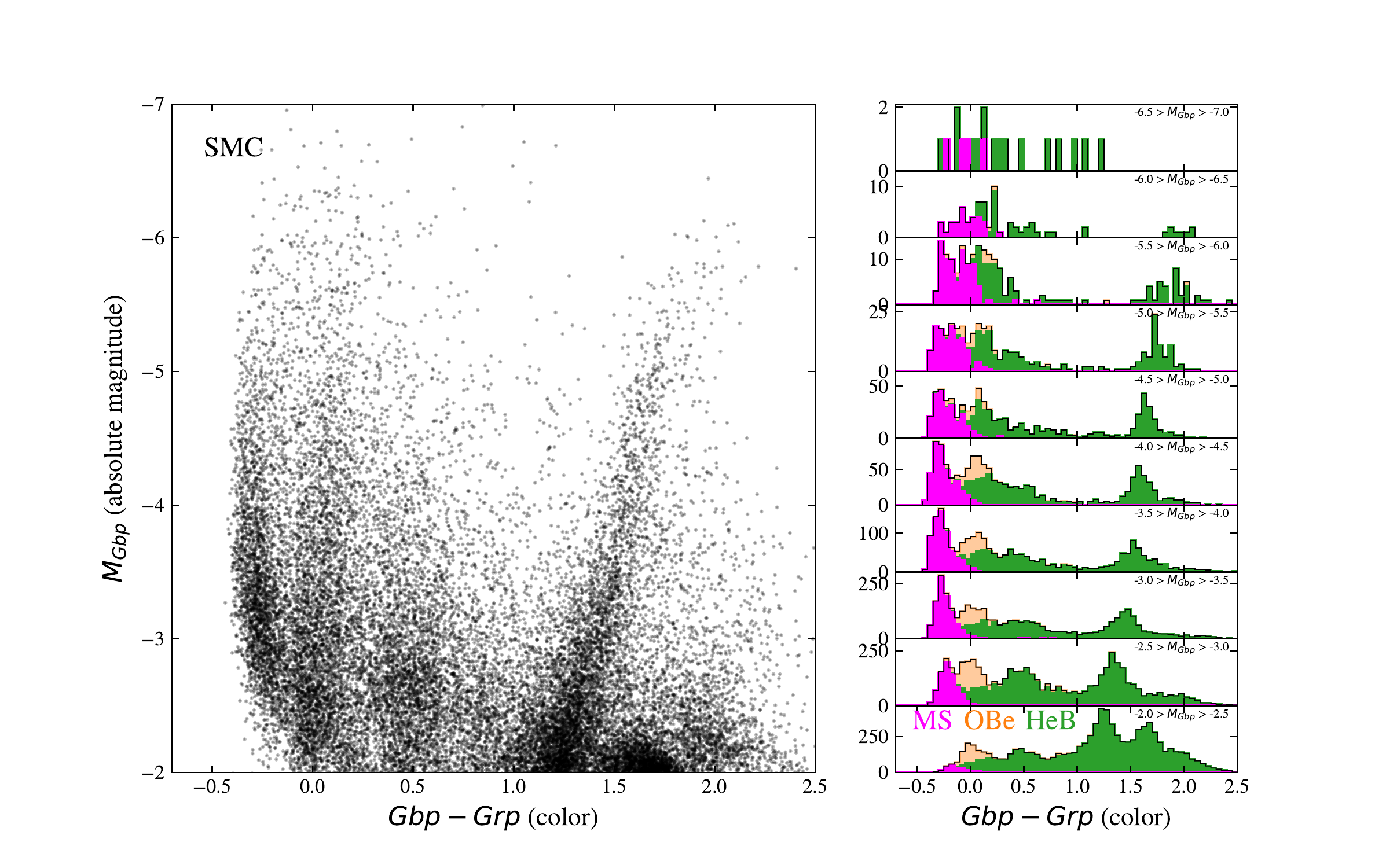}
\caption{\textit{Left:} color magnitude diagram of stars in the Small Magellanic Cloud (SMC), constructed with GAIA filters. Each black dot represents a source. The brightest sources are at the top and the bluest sources are on the left. 
We adopt a value of 18.91 for the distance modulus \citep{Hilditch05}.
\textit{Right:} stacked histograms of the $Gbp-Grp$ color in the absolute magnitude ranges that are written in the top right of each panel. Main-sequence (MS) stars are shown in magenta, OBe stars in light orange, and He-burning (HeB) stars in green. The black line shows the number distribution of all the sources combined.}
\label{fig:cmd}
\end{figure*}

\paragraph{Sextans A.}
For Sextans\,A, no deep enough GAIA and infrared data is available, and ground-based $UBVI$ data lack the required spatial resolution. Therefore, we use data obtained with the Hubble Space Telescope (HST). There is $F336W$ and $F439W$ photometry available in the catalog of \citep[][hereafter referred to as B12; obtained from \url{https://hla.stsci.edu/hlaview.html}]{Bianchi12}. These filters are centered on similar wavelengths as the $U$ and $B$ filters. 
B12 also registered photometry in the $F555$ and $F814W$ filters, which have central wavelengths similar to the GAIA $Gbp$ and $Grp$ filters. 
However, in the $F555$ and $F814W$ filters additional deeper observations exist \citep[][hereafter H06; from \url{http://astronomy.nmsu.edu/holtz/archival/html/lg.html}]{Holtzman06}.
Both B12 and H06 contain two HST fields in Sextans\,A, which cover about two thirds of the galaxy. To separate MS, OBe, and HeB stars in Sextans A, we cross-correlate B12 with H06 using TOPCAT \citep{Taylor05}, and check the result with the Aladin sky atlas \citep{Bonnarel00, Boch14}. We note that with a distance modulus $DM = 25.63$ \citep{Tammann11}, corresponding to a distance of 1.3\,Mpc, Sextans A is about twenty times further away that the Magellanic Clouds, which makes the cross-correlation a more difficult task.
Given a systematic coordinate offset in B12 and H06 data, we shift the B12 data before cross-matching. 
For data in frame 4 of B12, this shift is 1.321" in right ascension and 0.265" in declination. For their frame 5, the shift is 0.752" in right ascension and 1.087" in declination.

\paragraph{Holmberg I and Holmberg II.}
Our strategy to identify OBe stars in Holmberg\,I and Holmberg\,II is the same as for Sextans\,A, except that we employ data from the the Legacy ExtraGalactic UV Survey \cite[LEGUS;][from \url{https://archive.stsci.edu/prepds/legus/dataproducts-public.html}]{Sabbi18}, which is also obtained with HST. In LEGUS, the $F336W$, $F438W$, $F555W$, and $F814W$ filters are for simplicity referred to as the $U$, $B$, $V$, and $I$ filters, respectively. In this work we use the same nomenclature for LEGUS data. LEGUS $UBVI$ data cover most of Holmberg\,I and about a third of Holmberg\,II.



\begin{figure*}[ht]
\centering
\includegraphics[width=\linewidth]{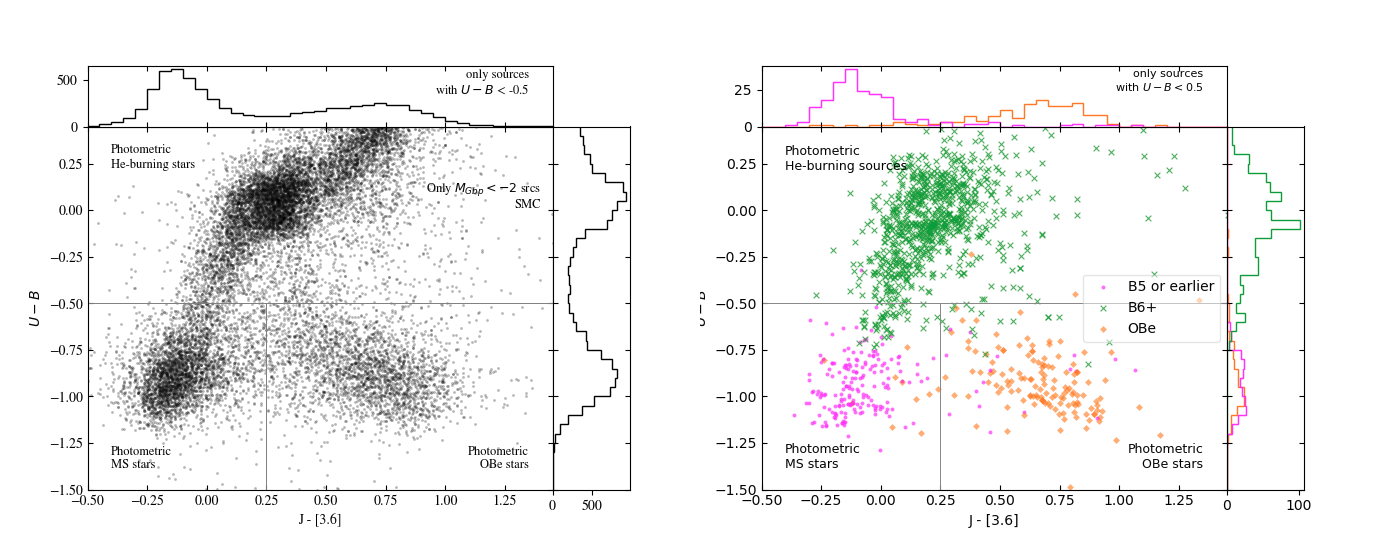}
\caption{\textit{Left:} color-color diagram of sources in our Small Magellanic Cloud (SMC) data set that are brighter than $M_\mathrm{Gbp} = -2$. Histograms are shown at the right panel and the top panel, where the top histogram includes only sources bluer than $U - B = -0.5$.  \textit{Right}: same as the left side, but showing sources from various spectroscopic studies \citep{Martayan07, Hunter08b, Dufton19, Bonanos10} that are colored for their spectral type. 
}
\label{fig:colocolo_3cut}
\end{figure*}

\section{Results \label{sec:results}}

\subsection{Small Magellanic Cloud}

We show the CMD of the sources in our SMC data set in the left panel of Fig.\,\ref{fig:cmd}.
At the right side there is the red supergiant branch, and on the left there are two bands of stars with blue colors, which we focus on in this work. 
These two blue bands lie at $Gbp - Grp \approx -0.25$ and at $Gbp - Grp \approx 0$. 

To investigate the nature of the sources in the two blue components, we plot the infrared $J - [3.6]$ color against the $U-B$ color in the left panel of Fig.\,\ref{fig:colocolo_3cut} for all sources in our SMC data set brighter than an absolute magnitude of $M_{Gbp} = -2$. 
Three distinct groups form, which can be interpreted with the aid of the location of stars with known spectral types in the right panel of Fig.\,\ref{fig:colocolo_3cut}. About 90\% of the sources in the bottom left group (i.e., with the bluest colors) that have known spectral types are O-type or early B-type stars. 
Such spectral types are typical for massive main sequence stars without decretion disks (hereafter referred to as `MS stars' for simplicity).
Similarly, about 90\% of sources in the group with a blue $U-B$ color but a red color in the infrared (photometric OBe stars) on the right side of Fig.\,\ref{fig:colocolo_3cut} are known to be Oe and early Be stars (see also histograms). Their location in the color-color diagram can be understood from their decretion disks emitting mainly at longer wavelengths, giving OBe sources a much redder color in the infrared \citep{Bonanos09, Bonanos10}, while their near-UV/blue optical $U-B$ color remains largely unaffected. Finally, in the top of the color-color diagrams there is a group of sources with $U - B > -0.5$. Most of these sources have spectral types of B6 or later (Fig.\,\ref{fig:colocolo_3cut}, right), implying that they have effective temperatures of $T_\mathrm{eff} \lesssim 15$\,kK \citep{Schootemeijer21}. Such effective temperatures explain the relatively red $U-B$ colors (compared to O and early B stars) and imply that the sources in this group are He-burning (HeB) objects, as the MS for massive SMC stars extends only up to $\sim$20\,kK even for extreme assumptions \citep{Schootemeijer19, Schootemeijer21}.

\begin{table*}[ht]
\caption{\label{tab:nbe}Number of OBe stars ($N_\mathrm{OBe}$) and OBe star fractions ($f_\mathrm{OBe}$) in the Large Magellanic Cloud (LMC), Small Magellanic Cloud (SMC), Holmberg\,I (HoI), Holmberg\,II (HoII) and Sextans\,A (SexA). The error given for $f_\mathrm{OBe}$ is the statistical error, which is a lower limit on the true error (see Appendix\,\ref{sec:appb_uncertainties}).} 
\small
\centering
\begin{tabular}{|l||l|l|l|l|l|l|l|l|l|l|}
\hline
Magnitude range & $N_\mathrm{OBe, \, LMC}$ & $f_\mathrm{OBe, \, LMC}$ & $N_\mathrm{OBe, \, SMC}$ & $f_\mathrm{OBe, \, SMC}$ & $N_\mathrm{OBe, \, HoI}$ & $f_\mathrm{OBe, \, HoI}$  & $N_\mathrm{OBe, \, HoII}$ & $f_\mathrm{OBe, \, HoII}$  & $N_\mathrm{OBe, \, SexA}$ & $f_\mathrm{OBe, \, SexA}$\\

\hline
\hline
$-5.5 < M_\mathrm{abs} < -5.0$ & 38 & 0.09 & 30 & 0.20 & 0 & 0 & 4 & 0.09 &1 & 0.20\\
$-5.0 < M_\mathrm{abs} < -4.5$ & 97 & 0.13 & 64 & 0.21 & 4 & 0.22 & 14 & 0.22 & 6 & 0.32\\
$-4.5 < M_\mathrm{abs} < -4.0$ & 267 & 0.20 & 166 & 0.30 & 13 & 0.30 & 32 & 0.22 & 11 & 0.34\\
$-4.0 < M_\mathrm{abs} < -3.5$ & 482 & 0.24 & 279 & 0.30 & 23 & 0.25 & 57 & 0.24 & 23 & 0.31\\
$-3.5 < M_\mathrm{abs} < -3.0$ & 819 & 0.24 & 494 & 0.33 & 43 & 0.27 & 137 & 0.30 & 22 & 0.21\\
$-3.0 < M_\mathrm{abs} < -2.5$ & 1169 & - & 787 & - & - & - & - & - & - & - \\ 
$-2.5 < M_\mathrm{abs} < -2.0$ & 1623 & - & 710 & - & - & - & - & - & - & - \\ 
\hline
$-5.0 < M_\mathrm{abs} < -3.0$ & 1665 & 0.22 & 1003 & 0.31 & 83 & 0.27 & 240 & 0.27 & 62 & 0.27\\
 &  & $\pm 0.01$ & & $\pm 0.01$ & & $\pm 0.03$ & & $\pm 0.02$ & & $\pm 0.03$\\
\hline
$-5.5 < M_\mathrm{abs} < -2.0$ & 4495 & - & 2530 & - & - & - & - & - & - & -\\
\hline
\end{tabular}
\end{table*}

The spectroscopic confirmation of the nature of the sources in the $U-B$ versus $J - [3.6]$ diagram implies that it is a very effective tool to identify the three types of blue massive stars in the SMC: MS (O/early B) stars, HeB (late B+) stars, and OBe stars.

At the right side of the CMD (Fig.\,\ref{fig:cmd}), we show histograms for MS, HeB, and OBe stars, as identified by their colors in Fig.\,\ref{fig:colocolo_3cut}. The bluest band at $Gbp-Grp \approx -0.25$ is almost exclusively made up of photometric MS stars.
The HeB sources occupy a broad range of colors, while OBe stars concentrate at $G_\mathrm{bp} - G_\mathrm{rp} \approx 0$ and give rise to a \textit{peak} in the number distribution there. This color difference of a quarter of a magnitude between MS and OBe stars was also found in previous work that used filters centered at similar wavelengths \citep{Martayan10, Milone18, Bodensteiner20}.
For sources brighter than $M_\mathrm{Gbp} =-5$ the number of OBe stars becomes very small, and the parallel feature starts to fade away. The total number of OBe stars shown in the CMD is large: slightly more than 2500 (Table\,\ref{tab:nbe}).

\subsection{The Large Magellanic Cloud, Holmberg I and II, and Sextans A}

We extend our analysis to four more dwarf galaxies, starting with the LMC.
As well as the SMC, the LMC is a satellite galaxy of the Milky Way, and it is slightly closer. However, with half the Solar metallicity \citep{Trundle07} it is more enriched in heavy elements than the SMC.
We can apply the same method because the LMC is covered by the data sets that we used for the SMC. 
We further describe this in 
Appendix\,\ref{sec:appb}, 
where we also discuss LMC's CMD (Fig.\,\ref{fig:cmd_lmc}) and the color-color diagram (Fig.\,\ref{fig:colocolo_3cut_lmc}).
We find as many as $\sim$4500 photometric OBe stars in the LMC that are brighter than $M_{Gbp} = -2$ (Table\,\ref{tab:nbe}).

Furthermore, we explore the dwarf galaxies Sextans\,A, Holmberg\,I (Ho\,I), and Holmberg\,II (Ho\,II), which all have a lower metallicity than the SMC and are much more distant.
Again, the more detailed description can be found in 
Appendix\,\ref{sec:appb}.
The CMDs of these galaxies are Figs.\,\ref{fig:cmd_sexa}, \ref{fig:cmd_hoi}, and \ref{fig:cmd_hoii}; the color-color diagrams are Figs.\,\ref{fig:colocolo_sexa}, \ref{fig:colocolo_hoi}, and \ref{fig:colocolo_hoii}.
Because we use different filters to identify OBe star disks in these dwarf galaxies, in Appendix\,\ref{sec:appb_id_obes} we provide multiple extra tests to demonstrate that with broadband photometry from HST, we can also tell apart O/early B stars, late B+ stars, and OBe stars. These tests involve stars with known spectral types, and narrowband H$\alpha$ photometry. Among these three dwarf galaxies, the most metal-poor one that we explore is Sextans\,A, with a metallicity $Z \approx 1/10$\,Z$_\odot$ \citep{Kaufer04, Garcia17}. It is a distance of about 1.3\,Mpc \citep{Tammann11}.
The final two dwarf galaxies, Ho\,I and Ho\,II, are at distances of $3-4$\,Mpc \citep{Sabbi18}. As such, these objects are beyond the Local Group, and belong to the M81 Group instead \citep{Karachentsev02}. According to their oxygen abundances,
Ho\,I and Ho\,II have metallicities of $\sim$1/6\,$Z_\odot$ and $\sim$1/7\,$Z_\odot$, respectively \citep{Croxall09, Bergemann21}.
In each of these three dwarf galaxies, which are smaller and less massive than the Magellanic Clouds, we identify of the order of one hundred OBe stars brighter than $M_\mathrm{abs} = -3$: 62 in Sextans\,A, 83 in Ho\,I, and 240 in Ho\,II (Table\,\ref{tab:nbe}).

\begin{figure}[ht]
\centering
\includegraphics[width=\linewidth]{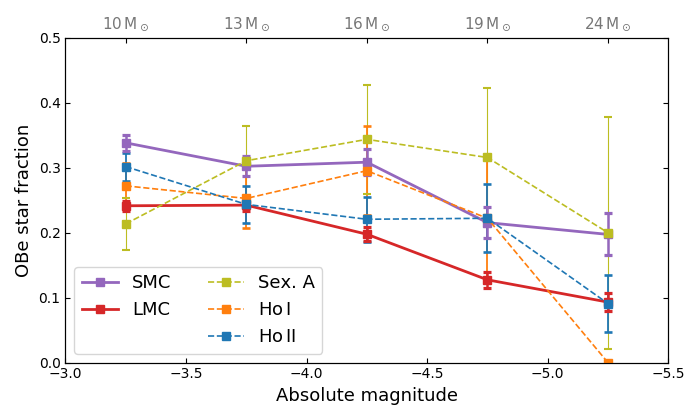}
\caption{Galaxy-wide OBe star fraction in the Magellanic Clouds, Sextans A, and Holmberg\,I and II, measured in absolute magnitude ranges $-3 > M_\mathrm{abs} > -3.5$, $-3.5 > M_\mathrm{abs} > -4$, and so on. The ticks at the top of the plot indicate the average evolutionary mass that has been inferred \citep{Schootemeijer21} for stars of spectral type B5 and earlier in each absolute magnitude bin in the SMC.}
\label{fig:fbe_both}
\end{figure}

\begin{figure}[ht]
\centering
\includegraphics[width=\linewidth]{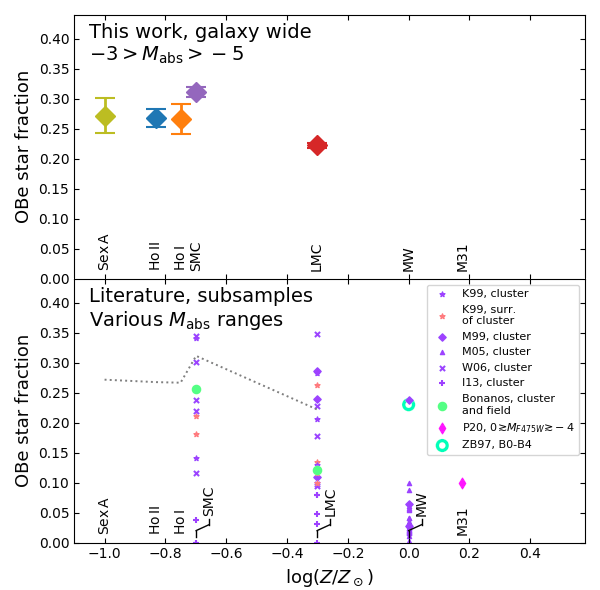}
\caption{Metallicity ($Z$) trends of the OBe star fraction. \textit{Top panel:} in the low-$Z$ regime, we show the galaxy-wide OBe star fraction calculated in this work in the absolute magnitude interval $-3 > M_\mathrm{abs} > -5$ with large diamonds.
\textit{Bottom panel:} literature studies of stellar sub-samples within different galaxies, which targeted cluster and/or field stars: K99 \citep{Keller99}, M99 \citep{Maeder99}, M05 \citep{McSwain05}, W06 \citep{Wisniewski06}, I13 \citep{Iqbal13}, Bonanos \citep{Bonanos09, Bonanos10}, P20 \citep{Peters20}, and ZB97 \citep{Zorec97}. We only show clusters with twenty or more OB+OBe stars, and we do not show multiple measurements for any cluster. 
In K99, M99, M05, W06, I13, and P20, narrow-band photometry was used to identify OBe stars. For a part of the sample in M99, OBe stars were identified with spectroscopy. Bonanos used infrared excess of stars with known spectral types to identify OBe stars. ZB97 used a sample of stars with known spectral types. For reference we show the results of this work again with a dotted line. The metallicities in Holmberg\,I and II (Ho\,I and II) are estimated using oxygen abundances \citep{Croxall09}.
The two highest-metallicity galaxies are our own Milky Way (MW) and Andromeda (M31). For $Z_\mathrm{MW}$ we adopt the Solar metallicity, and we adopt $Z_\mathrm{M31} = 1.5$\,Z$_\odot$ from P20.
}
\label{fig:Z_trends}
\end{figure}

\begin{figure}[ht]
\centering
\includegraphics[width=\linewidth]{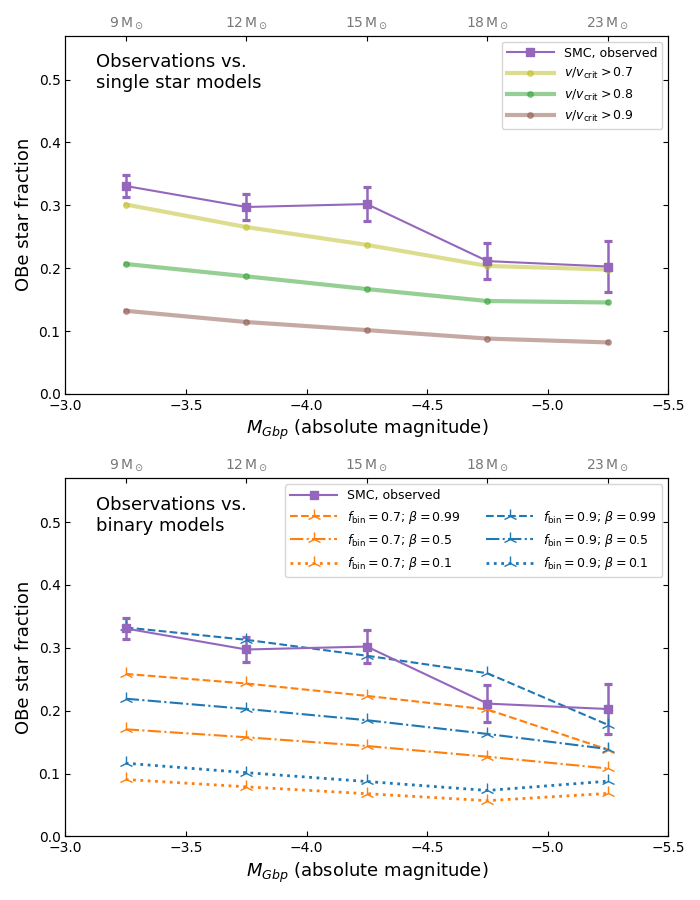}
\caption{
Comparison of the observed absolute magnitude dependent SMC OBe star fraction with theory. At the top of each panel we show indicative masses, as in Fig.\,\ref{fig:fbe_both} \textit{Top:} OBe star fraction in a theoretical single star population, for which we explore three different minimum $v/v_\mathrm{crit}$ values that stars require to manifest themselves as OBe stars, $v/v_\mathrm{crit}$ being the fraction of the critical rotation velocity.
\textit{Bottom:} OBe star fraction in theoretical binary populations. With blue and orange lines, we show theoretical predictions derived from ComBinE simulations, for different combinations of the mass transfer efficiency $\beta$ and binary fraction $f_\mathrm{bin}$. In the theoretical populations, single stars are assumed to not contribute to producing OBe stars. For each theoretical binary population, we assume that all stars with $v/v_\mathrm{crit} > 0.7$ are OBe stars. In the Appendix, we provide a similar figure, but where $v/v_\mathrm{crit} > 0.8$ and $v/v_\mathrm{crit} > 0.9$ are set as a required rotation velocities for OBe stars in the binary populations (Fig.\,\ref{fig:fbe_smc_vs_theory_vvcbin0809}).}
\label{fig:fbe_smc_vs_theory}
\end{figure}

\subsection{Magnitude-, mass-, and metallicity-dependent OBe star fractions}
To calculate the OBe star fraction in different absolute magnitude bins, we divide the number of OBe stars, $N_\mathrm{OBe}$, by the total number of H-burning stars, $N_\mathrm{OBe} + N_\mathrm{MS}$ (selected by the color cuts in Figs.\,\ref{fig:colocolo_3cut}, \ref{fig:colocolo_3cut_lmc}, \ref{fig:colocolo_sexa}, \ref{fig:colocolo_hoi}, and \ref{fig:colocolo_hoii}). 
When constructing Fig.\,\ref{fig:fbe_both}, we assumed that the disk adds no significant flux to the $Gbp$, $V$, or $F555W$ filter (which is what we find in the Appendix\,\ref{sec:appb_uncertainties}).
MS sources in the SMC that are dimmer than $M_{abs} \approx -3$ are not bright enough to be observed in Spitzer's [3.6] filter. For the LMC and Sextans A, the completeness fractions of OBe stars and non-OBe MS stars also start to significantly deviate at a similar absolute magnitude (Appendix\,\ref{sec:appb_completeness}). Therefore, we cannot accurately calculate the OBe star fraction at $M_{abs} \gtrsim -3$.
At the bright end ($M_\mathrm{abs} < -5.5$), we notice that the fraction of Wolf-Rayet and B[e] stars posing as OBe stars becomes significant, in particular in the LMC (Appendix\,\ref{sec:appb_pollution}). Therefore, we only display the absolute magnitude range $-3 > M_\mathrm{abs} > -5.5$ in Fig.\,\ref{fig:fbe_both}. The values we find for $N_\mathrm{OBe}$ and $f_\mathrm{OBe}$ are listed in Table\,\ref{tab:nbe}. 
At the top of Fig.\,\ref{fig:fbe_both}, we write in each absolute magnitude interval the average evolutionary mass that has been determined previously \citep{Schootemeijer21} for SMC stars of type B5 and earlier (see Appendix\,\ref{sec:appc}). 
The obtained masses indicate that the sources shown in Fig.\,\ref{fig:fbe_both} are typically massive stars with masses up to 25\,M$_\odot$.
Given the spread in evolutionary masses and potential biases, these masses are meant to serve as an indication.

We find that the OBe star fraction is higher in the SMC than in the LMC. 
In the LMC, $f_\mathrm{OBe} \approx 0.25$ at 10M$_\odot$, but then it decreases quickly for $M\gtrsim15$M$_\odot$. In the SMC, $f_\mathrm{OBe} \approx 0.35$ around 10\,M$_\odot$, and it decreases to $\sim$0.20 around 20M$_\odot$. The relatively large difference in OBe star fraction between the SMC and LMC towards $M_{Gbp} = -5$ could be the result of stellar winds (which are stronger at higher metallicity) spinning down the bright stars more strongly in the LMC.
However, for disk angular momentum loss of OBe stars \citep{Okazaki01}, the metallicity-dependence is not as well understood.
With further decreasing metallicity in Ho\,I, Ho\,II, and Sextans\,A, we find OBe star fractions that are comparable to what we find in the SMC.
This becomes more apparent in Fig.\,\ref{fig:Z_trends}, where we plot our galaxy-wide OBe star fractions against metallicity.
In all four galaxies in the regime $1/10 \leq Z/Z_\odot \leq 1/5$, about 30\% of the stars in the considered magnitude range are OBe stars. This implies that extremely rapid rotation is commonplace in metal-poor environments.

In the SMC and LMC, we can discuss our results in the context of literature values reported for stellar sub-samples. However, we caution that a direct comparison is inappropriate because of different sample selection, and time-evolution of cluster OBe star fractions. Indeed, as a result of this Fig.\,\ref{fig:Z_trends} shows strong variation from cluster to cluster within individual galaxies. Also, investigated clusters are often older than 40\,Myr, which is roughly the lifetime of an 8\,M$_\odot$ star. Therefore, they contain mainly B and Be stars below 8\,M$_\odot$ rather than massive stars, which are the topic of this work. The LMC and SMC OBe star fractions that we find are above the average literature values for cluster studies. They are also higher than results for sub-samples of stars with known spectral types (green circles), possibly because spectral type catalogs can biased towards the brighter stars, for which we find lower OBe star fractions (Fig.\,\ref{fig:fbe_both}).

\subsection{Implications for OBe star formation}
It is currently debated whether OBe stars are mainly produced by single star evolution \citep{vanBever97} or through binary evolution \citep{Schootemeijer18b, Klement19, Bodensteiner20b} where mass transfer from one star to the other leads to spin-up and the production of an OBe star. In this light, we briefly discuss the high value of $f_\mathrm{OBe} \approx 0.3$ in the SMC from a  theoretical perspective.
Previous work, which used the rotational velocity distribution of \cite{Dufton13} as the initial distribution, has shown that the single star channel struggles to produce the (high) observed fraction of OBe stars in SMC star cluster NGC\,330 \citep{Hastings20}. 
Here we redo the analysis of \cite{Hastings20} with the difference that for the theoretical population we assume that it undergoes constant star formation (CSF) instead of it being coeval -- for details, see Appendix\,\ref{sec:app_ss}.
Fig.\,\ref{fig:fbe_smc_vs_theory} shows that for CSF, the single star population can match the observed OBe star fraction as long as $v/v_\mathrm{crit} \gtrsim 0.7$ ($v_\mathrm{crit}$ being the breakup velocity) is sufficient for stars to manifest themselves as OBe stars. However, if stars have to rotate at $v/v_\mathrm{crit} \gtrsim 0.8 - 0.9$ to display the OBe phenomenon \citep{Townsend04, Fremat05}, the single star channel cannot explain the high observed OBe star fractions in the SMC in our simulations. 

On the other hand, the binary channel also struggles to explain the observed OBe star fraction. In the SMC (and Ho\,I, Ho\,II, and Sextans\,A), the OBe star fraction is so high that it touches the stringent upper limit of $f_\mathrm{OBe} \lesssim 1 / 3$ at and below the turnoff in coeval star clusters \citep{Hastings21}. However, for a galaxy-wide population undergoing CSF, efficient mass transfer can help to increase the OBe star fraction shown in Fig.\,\ref{fig:fbe_both} because it can turn lower-mass stars -- which are favored by the Initial Mass Function -- into brighter OBe stars. To test this, we simulate three SMC binary populations with the rapid population synthesis code ComBinE \citep{Kruckow18}, varying the mass transfer efficiency (Fig.\,\ref{fig:fbe_smc_vs_theory}). We elaborate on these simulations in Appendix\,\ref{sec:app_bin}.
In the theoretical population, for inefficient to moderately efficient mass transfer ($\beta \leq 0.5$), the binary channel cannot explain an OBe star fraction as high as 0.3, even for 90\% of sources being interacting binaries.
For conservative mass transfer, however, it does become possible to reproduce $f_\mathrm{OBe} \approx 0.3$. 
As is the case for the single star simulations, our binary simulations do a good job at explaining the behaviour of the OBe star fraction as a function of absolute magnitude, which strengthens our results.
In principle, a scenario in which stars in binaries spin up to become OBe stars before mass exchange takes place could help to explain the high observed OBe star fractions. However, recent work found no clear evidence for any MS companion in a sample of over 250 Galactic Be stars \citep{Bodensteiner20b}, suggesting that OBe star formation before interaction in binaries rarely takes place.

\section{Discussion \& conclusions}

We found that in color-color diagrams of metal-poor dwarf galaxies, main sequence stars, helium-burning stars, and OBe stars form well-separated groups of sources. We took advantage of this to identify OBe stars based on broad-band photometry, and as such 
we could apply this method to
far-away galaxies. We analyzed the Milky Way satellite dwarf galaxies LMC (50\,kpc; 1/2\,Z$_\odot$) and SMC (60\,kpc; 1/5\,Z$_\odot$), and three of such distant galaxies: Sextans\,A (1.3\,Mpc away, at the border of the Local Group; 1/10\,Z$_\odot$), and in the M81 group Holmberg\,I (3.8\,Mpc; 1/6\,Z$_\odot$) and Holmberg\,II (3.2\,Mpc; 1/7\,Z$_\odot$).
Using archival photometric data, we were able to find of the order of a hundred to a few thousand OBe stars in each galaxy.
This allowed us to calculate galaxy-wide OBe star fractions.

We found that the OBe star fraction increases with decreasing metallicity between the LMC and the SMC, and stays high -- around 30\% -- in Sextans\,A, Holmberg\,I, and Holmberg\,II. In other words, extremely rapidly rotating massive stars are common in low-metallicity dwarf galaxies. These high OBe star fractions shed light on their formation channel. Comparing to SMC population synthesis models, we found that the single star channel can only reproduce the high observed OBe star fractions if significantly sub-critical rotation suffices to display the OBe phenomenon. For the binary channel to produce enough bright OBe stars, the binary fraction has to be close to unity and -- at the same time -- mass transfer has to be efficient. 

Our work provides galaxy-wide OBe star fraction measurements in the Magellanic Clouds, and we extended OBe star fraction measurements to sub-SMC metallicity and beyond the Local Group. Within the now-accessible volume, our method can be applied to even more metal-poor galaxies to further test the behavior of the OBe star fraction at low $Z$. There are a number of such galaxies that are currently star-forming \citep{Garcia21}, but would require near-UV observations in addition to the existing optical photometry \citep[e.g., ANGST --][]{Dalcanton09}. Examples of promising targets are Sextans B \citep[1/10\,Z$_\odot$ to 1/15\,Z$_\odot$:][]{Kniazev05}, Leo A \citep[$\sim$1/20\,Z$_\odot$:][]{vanZee06}, and SagDIG \citep[$\sim$ 1/20\,Z$_\odot$:][]{Saviane02}. Fortunately, such efforts are currently being undertaken \citep[e.g.,][]{Sabbi20}.
In the galaxies up to the distance of $\sim$4\,Mpc reached here, OBe stars in the range $-3 > M_\mathrm{abs} > -5$ did not hit HST's detection limit. Therefore, even more distant galaxies could be studied, as long as the data quality remains high and color-color diagrams similar to the ones verified in this work are used to identify OBe stars.
Exploiting such galaxies with sub-Sextans\,A metallicities would allow to get even closer to measuring the incidence of extremely rapid rotation among massive stars in environments that resemble the primordial universe.

\begin{acknowledgements}
We thank the anonymous referee for their careful reading of the manuscript and for useful comments.
M.G. gratefully acknowledges support by grants PID2019-105552RB-C41 and MDM-2017-0737
Unidad de Excelencia "Mar\'{\i}a de Maeztu"-Centro de Astrobiolog\'{\i}a (CSIC-INTA), funded
by the Spanish Ministry of Science and Innovation/State Agency of Research MCIN/AEI/10.13039/501100011033.
\end{acknowledgements}

\typeout{}
\bibliography{sample631}{}

\begin{thebibliography}{104}
\expandafter\ifx\csname natexlab\endcsname\relax\def\natexlab#1{#1}\fi

\bibitem[{{Aguilera-Dena} {et~al.}(2018){Aguilera-Dena}, {Langer}, {Moriya}, \&
  {Schootemeijer}}]{Aguilera18}
{Aguilera-Dena}, D.~R., {Langer}, N., {Moriya}, T.~J., \& {Schootemeijer}, A.
  2018, \apj, 858, 115

\bibitem[{{Arenou} {et~al.}(2018){Arenou}, {Luri}, {Babusiaux}, {Fabricius},
  {Helmi}, {Muraveva}, {Robin}, {Spoto}, {Vallenari}, {Antoja},
  {Cantat-Gaudin}, {Jordi}, {Leclerc}, {Reyl{\'e}}, {Romero-G{\'o}mez}, {Shih},
  {Soria}, {Barache}, {Bossini}, {Bragaglia}, {Breddels}, {Fabrizio},
  {Lambert}, 2~{Marrese}, {Massari}, {Moitinho}, {Robichon}, {Ruiz-Dern},
  {Sordo}, {Veljanoski}, {Eyer}, {Jasniewicz}, {Pancino}, {Soubiran}, {Spagna},
  {Tanga}, {Turon}, \& {Zurbach}}]{Arenou18}
{Arenou}, F., {Luri}, X., {Babusiaux}, C., {et~al.} 2018, \aap, 616, A17

\bibitem[{{Bergemann} {et~al.}(2021){Bergemann}, {Hoppe}, {Semenova},
  {Carlsson}, {Yakovleva}, {Voronov}, {Bautista}, {Nemer}, {Belyaev},
  {Leenaarts}, {Mashonkina}, {Reiners}, \& {Ellwarth}}]{Bergemann21}
{Bergemann}, M., {Hoppe}, R., {Semenova}, E., {et~al.} 2021, \mnras, 508, 2236

\bibitem[{{Bianchi} {et~al.}(2012){Bianchi}, {Efremova}, {Hodge}, {Massey}, \&
  {Olsen}}]{Bianchi12}
{Bianchi}, L., {Efremova}, B., {Hodge}, P., {Massey}, P., \& {Olsen}, K.~A.~G.
  2012, \aj, 143, 74

\bibitem[{{Boch} \& {Fernique}(2014)}]{Boch14}
{Boch}, T. \& {Fernique}, P. 2014, in Astronomical Society of the Pacific
  Conference Series, Vol. 485, Astronomical Data Analysis Software and Systems
  XXIII, ed. N.~{Manset} \& P.~{Forshay}, 277

\bibitem[{{Boch} {et~al.}(2012){Boch}, {Pineau}, \& {Derriere}}]{Boch12}
{Boch}, T., {Pineau}, F., \& {Derriere}, S. 2012, in Astronomical Society of
  the Pacific Conference Series, Vol. 461, Astronomical Data Analysis Software
  and Systems XXI, ed. P.~{Ballester}, D.~{Egret}, \& N.~P.~F. {Lorente}, 291

\bibitem[{{Bodensteiner} {et~al.}(2020{\natexlab{a}}){Bodensteiner}, {Sana},
  {Mahy}, {Patrick}, {de Koter}, {de Mink}, {Evans}, {G{\"o}tberg}, {Langer},
  {Lennon}, {Schneider}, \& {Tramper}}]{Bodensteiner20}
{Bodensteiner}, J., {Sana}, H., {Mahy}, L., {et~al.} 2020{\natexlab{a}}, \aap,
  634, A51

\bibitem[{{Bodensteiner} {et~al.}(2020{\natexlab{b}}){Bodensteiner}, {Shenar},
  \& {Sana}}]{Bodensteiner20b}
{Bodensteiner}, J., {Shenar}, T., \& {Sana}, H. 2020{\natexlab{b}}, \aap, 641,
  A42

\bibitem[{{Bonanos} {et~al.}(2010){Bonanos}, {Lennon}, {K{\"o}hlinger}, {van
  Loon}, {Massa}, {Sewilo}, {Evans}, {Panagia}, {Babler}, {Block}, {Bracker},
  {Engelbracht}, {Gordon}, {Hora}, {Indebetouw}, {Meade}, {Meixner}, {Misselt},
  {Robitaille}, {Shiao}, \& {Whitney}}]{Bonanos10}
{Bonanos}, A.~Z., {Lennon}, D.~J., {K{\"o}hlinger}, F., {et~al.} 2010, \aj,
  140, 416

\bibitem[{{Bonanos} {et~al.}(2009){Bonanos}, {Massa}, {Sewilo}, {Lennon},
  {Panagia}, {Smith}, {Meixner}, {Babler}, {Bracker}, {Meade}, {Gordon},
  {Hora}, {Indebetouw}, \& {Whitney}}]{Bonanos09}
{Bonanos}, A.~Z., {Massa}, D.~L., {Sewilo}, M., {et~al.} 2009, \aj, 138, 1003

\bibitem[{{Bonnarel} {et~al.}(2000){Bonnarel}, {Fernique}, {Bienaym{\'e}},
  {Egret}, {Genova}, {Louys}, {Ochsenbein}, {Wenger}, \&
  {Bartlett}}]{Bonnarel00}
{Bonnarel}, F., {Fernique}, P., {Bienaym{\'e}}, O., {et~al.} 2000, \aaps, 143,
  33

\bibitem[{{Brott} {et~al.}(2011){Brott}, {de Mink}, {Cantiello}, {Langer}, {de
  Koter}, {Evans}, {Hunter}, {Trundle}, \& {Vink}}]{Brott11}
{Brott}, I., {de Mink}, S.~E., {Cantiello}, M., {et~al.} 2011, \aap, 530, A115

\bibitem[{{Choi} {et~al.}(2016){Choi}, {Dotter}, {Conroy}, {Cantiello},
  {Paxton}, \& {Johnson}}]{Choi16}
{Choi}, J., {Dotter}, A., {Conroy}, C., {et~al.} 2016, \apj, 823, 102

\bibitem[{{Corsaro} {et~al.}(2017){Corsaro}, {Lee}, {Garc{\'\i}a},
  {Hennebelle}, {Mathur}, {Beck}, {Mathis}, {Stello}, \& {Bouvier}}]{Corsaro17}
{Corsaro}, E., {Lee}, Y.-N., {Garc{\'\i}a}, R.~A., {et~al.} 2017, Nature
  Astronomy, 1, 0064

\bibitem[{{Croxall} {et~al.}(2009){Croxall}, {van Zee}, {Lee}, {Skillman},
  {Lee}, {C{\^o}t{\'e}}, {Kennicutt}, \& {Miller}}]{Croxall09}
{Croxall}, K.~V., {van Zee}, L., {Lee}, H., {et~al.} 2009, \apj, 705, 723

\bibitem[{{Dalcanton} {et~al.}(2009){Dalcanton}, {Williams}, {Seth}, {Dolphin},
  {Holtzman}, {Rosema}, {Skillman}, {Cole}, {Girardi}, {Gogarten},
  {Karachentsev}, {Olsen}, {Weisz}, {Christensen}, {Freeman}, {Gilbert},
  {Gallart}, {Harris}, {Hodge}, {de Jong}, {Karachentseva}, {Mateo}, {Stetson},
  {Tavarez}, {Zaritsky}, {Governato}, \& {Quinn}}]{Dalcanton09}
{Dalcanton}, J.~J., {Williams}, B.~F., {Seth}, A.~C., {et~al.} 2009, \apjs,
  183, 67

\bibitem[{{Dallas} {et~al.}(2022){Dallas}, {Oey}, \& {Castro}}]{Dallas22}
{Dallas}, M.~M., {Oey}, M.~S., \& {Castro}, N. 2022, arXiv e-prints,
  arXiv:2208.10408

\bibitem[{{Dayal} \& {Ferrara}(2018)}]{Dayal18}
{Dayal}, P. \& {Ferrara}, A. 2018, \physrep, 780, 1

\bibitem[{{de Mink} \& {Mandel}(2016)}]{deMink16}
{de Mink}, S.~E. \& {Mandel}, I. 2016, \mnras, 460, 3545

\bibitem[{{Dohm-Palmer} {et~al.}(2002){Dohm-Palmer}, {Skillman}, {Mateo},
  {Saha}, {Dolphin}, {Tolstoy}, {Gallagher}, \& {Cole}}]{Dohm-Palmer02}
{Dohm-Palmer}, R.~C., {Skillman}, E.~D., {Mateo}, M., {et~al.} 2002, \aj, 123,
  813

\bibitem[{{Dotter}(2016)}]{Dotter16}
{Dotter}, A. 2016, \apjs, 222, 8

\bibitem[{{Dufton} {et~al.}(2019){Dufton}, {Evans}, {Hunter}, {Lennon}, \&
  {Schneider}}]{Dufton19}
{Dufton}, P.~L., {Evans}, C.~J., {Hunter}, I., {Lennon}, D.~J., \& {Schneider},
  F.~R.~N. 2019, \aap, 626, A50

\bibitem[{{Dufton} {et~al.}(2013){Dufton}, {Langer}, {Dunstall}, {Evans},
  {Brott}, {de Mink}, {Howarth}, {Kennedy}, {McEvoy}, {Potter},
  {Ram{\'\i}rez-Agudelo}, {Sana}, {Sim{\'o}n-D{\'\i}az}, {Taylor}, \&
  {Vink}}]{Dufton13}
{Dufton}, P.~L., {Langer}, N., {Dunstall}, P.~R., {et~al.} 2013, \aap, 550,
  A109

\bibitem[{{Ekstr{\"o}m} {et~al.}(2008){Ekstr{\"o}m}, {Meynet}, {Maeder}, \&
  {Barblan}}]{Ekstrom08}
{Ekstr{\"o}m}, S., {Meynet}, G., {Maeder}, A., \& {Barblan}, F. 2008, \aap,
  478, 467

\bibitem[{{Erb} {et~al.}(2010){Erb}, {Pettini}, {Shapley}, {Steidel}, {Law}, \&
  {Reddy}}]{Erb10}
{Erb}, D.~K., {Pettini}, M., {Shapley}, A.~E., {et~al.} 2010, \apj, 719, 1168

\bibitem[{{Espinosa Lara} \& {Rieutord}(2011)}]{EspinosaLara11}
{Espinosa Lara}, F. \& {Rieutord}, M. 2011, \aap, 533, A43

\bibitem[{{Evans} {et~al.}(2019){Evans}, {Castro}, {Gonzalez}, {Garcia},
  {Bastian}, {Cioni}, {Clark}, {Davies}, {Ferguson}, {Kamann}, {Lennon},
  {Patrick}, {Vink}, \& {Weisz}}]{Evans19}
{Evans}, C.~J., {Castro}, N., {Gonzalez}, O.~A., {et~al.} 2019, \aap, 622, A129

\bibitem[{{Evans} {et~al.}(2004){Evans}, {Howarth}, {Irwin}, {Burnley}, \&
  {Harries}}]{Evans04}
{Evans}, C.~J., {Howarth}, I.~D., {Irwin}, M.~J., {Burnley}, A.~W., \&
  {Harries}, T.~J. 2004, \mnras, 353, 601

\bibitem[{{Evans} {et~al.}(2006){Evans}, {Lennon}, {Smartt}, \&
  {Trundle}}]{Evans06}
{Evans}, C.~J., {Lennon}, D.~J., {Smartt}, S.~J., \& {Trundle}, C. 2006, \aap,
  456, 623

\bibitem[{{Fr{\'e}mat} {et~al.}(2005){Fr{\'e}mat}, {Zorec}, {Hubert}, \&
  {Floquet}}]{Fremat05}
{Fr{\'e}mat}, Y., {Zorec}, J., {Hubert}, A.~M., \& {Floquet}, M. 2005, \aap,
  440, 305

\bibitem[{{Gaia Collaboration} {et~al.}(2021){Gaia Collaboration}, {Brown},
  {Vallenari}, {Prusti}, {de Bruijne}, {Babusiaux}, {Biermann}, {Creevey},
  {Evans}, {Eyer}, {Hutton}, {Jansen}, {Jordi}, {Klioner}, {Lammers},
  {Lindegren}, {Luri}, {Mignard}, {Panem}, {Pourbaix}, {Randich}, {Sartoretti},
  {Soubiran}, {Walton}, {Arenou}, {Bailer-Jones}, {Bastian}, {Cropper},
  {Drimmel}, {Katz}, {Lattanzi}, {van Leeuwen}, {Bakker}, {Cacciari},
  {Casta{\~n}eda}, {De Angeli}, {Ducourant}, {Fabricius}, {Fouesneau},
  {Fr{\'e}mat}, {Guerra}, {Guerrier}, {Guiraud}, {Jean-Antoine Piccolo},
  {Masana}, {Messineo}, {Mowlavi}, {Nicolas}, {Nienartowicz}, {Pailler},
  {Panuzzo}, {Riclet}, {Roux}, {Seabroke}, {Sordo}, {Tanga}, {Th{\'e}venin},
  {Gracia-Abril}, {Portell}, {Teyssier}, {Altmann}, {Andrae}, {Bellas-Velidis},
  {Benson}, {Berthier}, {Blomme}, {Brugaletta}, {Burgess}, {Busso}, {Carry},
  {Cellino}, {Cheek}, {Clementini}, {Damerdji}, {Davidson}, {Delchambre},
  {Dell'Oro}, {Fern{\'a}ndez-Hern{\'a}ndez}, {Galluccio}, {Garc{\'\i}a-Lario},
  {Garcia-Reinaldos}, {Gonz{\'a}lez-N{\'u}{\~n}ez}, {Gosset}, {Haigron},
  {Halbwachs}, {Hambly}, {Harrison}, {Hatzidimitriou}, {Heiter},
  {Hern{\'a}ndez}, {Hestroffer}, {Hodgkin}, {Holl}, {Jan{\ss}en}, {Jevardat de
  Fombelle}, {Jordan}, {Krone-Martins}, {Lanzafame}, {L{\"o}ffler}, {Lorca},
  {Manteiga}, {Marchal}, {Marrese}, {Moitinho}, {Mora}, {Muinonen}, {Osborne},
  {Pancino}, {Pauwels}, {Petit}, {Recio-Blanco}, {Richards}, {Riello},
  {Rimoldini}, {Robin}, {Roegiers}, {Rybizki}, {Sarro}, {Siopis}, {Smith},
  {Sozzetti}, {Ulla}, {Utrilla}, {van Leeuwen}, {van Reeven}, {Abbas}, {Abreu
  Aramburu}, {Accart}, {Aerts}, {Aguado}, {Ajaj}, {Altavilla}, {{\'A}lvarez},
  {{\'A}lvarez Cid-Fuentes}, {Alves}, {Anderson}, {Anglada Varela}, {Antoja},
  {Audard}, {Baines}, {Baker}, {Balaguer-N{\'u}{\~n}ez}, {Balbinot}, {Balog},
  {Barache}, {Barbato}, {Barros}, {Barstow}, {Bartolom{\'e}}, {Bassilana},
  {Bauchet}, {Baudesson-Stella}, {Becciani}, {Bellazzini}, {Bernet}, {Bertone},
  {Bianchi}, {Blanco-Cuaresma}, {Boch}, {Bombrun}, {Bossini}, {Bouquillon},
  {Bragaglia}, {Bramante}, {Breedt}, {Bressan}, {Brouillet}, {Bucciarelli},
  {Burlacu}, {Busonero}, {Butkevich}, {Buzzi}, {Caffau}, {Cancelliere},
  {C{\'a}novas}, {Cantat-Gaudin}, {Carballo}, {Carlucci}, {Carnerero},
  {Carrasco}, {Casamiquela}, {Castellani}, {Castro-Ginard}, {Castro Sampol},
  {Chaoul}, {Charlot}, {Chemin}, {Chiavassa}, {Cioni}, {Comoretto}, {Cooper},
  {Cornez}, {Cowell}, {Crifo}, {Crosta}, {Crowley}, {Dafonte}, {Dapergolas},
  {David}, {David}, {de Laverny}, {De Luise}, {De March}, {De Ridder}, {de
  Souza}, {de Teodoro}, {de Torres}, {del Peloso}, {del Pozo}, {Delbo},
  {Delgado}, {Delgado}, {Delisle}, {Di Matteo}, {Diakite}, {Diener},
  {Distefano}, {Dolding}, {Eappachen}, {Edvardsson}, {Enke}, {Esquej}, {Fabre},
  {Fabrizio}, {Faigler}, {Fedorets}, {Fernique}, {Fienga}, {Figueras},
  {Fouron}, {Fragkoudi}, {Fraile}, {Franke}, {Gai}, {Garabato},
  {Garcia-Gutierrez}, {Garc{\'\i}a-Torres}, {Garofalo}, {Gavras}, {Gerlach},
  {Geyer}, {Giacobbe}, {Gilmore}, {Girona}, {Giuffrida}, {Gomel}, {Gomez},
  {Gonzalez-Santamaria}, {Gonz{\'a}lez-Vidal}, {Granvik},
  {Guti{\'e}rrez-S{\'a}nchez}, {Guy}, {Hauser}, {Haywood}, {Helmi}, {Hidalgo},
  {Hilger}, {H{\l}adczuk}, {Hobbs}, {Holland}, {Huckle}, {Jasniewicz},
  {Jonker}, {Juaristi Campillo}, {Julbe}, {Karbevska}, {Kervella}, {Khanna},
  {Kochoska}, {Kontizas}, {Kordopatis}, {Korn}, {Kostrzewa-Rutkowska},
  {Kruszy{\'n}ska}, {Lambert}, {Lanza}, {Lasne}, {Le Campion}, {Le Fustec},
  {Lebreton}, {Lebzelter}, {Leccia}, {Leclerc}, {Lecoeur-Taibi}, {Liao},
  {Licata}, {Lindstr{\o}m}, {Lister}, {Livanou}, {Lobel}, {Madrero Pardo},
  {Managau}, {Mann}, {Marchant}, {Marconi}, {Marcos Santos}, {Marinoni},
  {Marocco}, {Marshall}, {Martin Polo}, {Mart{\'\i}n-Fleitas}, {Masip},
  {Massari}, {Mastrobuono-Battisti}, {Mazeh}, {McMillan}, {Messina},
  {Michalik}, {Millar}, {Mints}, {Molina}, {Molinaro}, {Moln{\'a}r},
  {Montegriffo}, {Mor}, {Morbidelli}, {Morel}, {Morris}, {Mulone}, {Munoz},
  {Muraveva}, {Murphy}, {Musella}, {Noval}, {Ord{\'e}novic}, {Orr{\`u}},
  {Osinde}, {Pagani}, {Pagano}, {Palaversa}, {Palicio}, {Panahi}, {Pawlak},
  {Pe{\~n}alosa Esteller}, {Penttil{\"a}}, {Piersimoni}, {Pineau}, {Plachy},
  {Plum}, {Poggio}, {Poretti}, {Poujoulet}, {Pr{\v{s}}a}, {Pulone}, {Racero},
  {Ragaini}, {Rainer}, {Raiteri}, {Rambaux}, {Ramos}, {Ramos-Lerate}, {Re
  Fiorentin}, {Regibo}, {Reyl{\'e}}, {Ripepi}, {Riva}, {Rixon}, {Robichon},
  {Robin}, {Roelens}, {Rohrbasser}, {Romero-G{\'o}mez}, {Rowell}, {Royer},
  {Rybicki}, {Sadowski}, {Sagrist{\`a} Sell{\'e}s}, {Sahlmann}, {Salgado},
  {Salguero}, {Samaras}, {Sanchez Gimenez}, {Sanna}, {Santove{\~n}a},
  {Sarasso}, {Schultheis}, {Sciacca}, {Segol}, {Segovia}, {S{\'e}gransan},
  {Semeux}, {Shahaf}, {Siddiqui}, {Siebert}, {Siltala}, {Slezak}, {Smart},
  {Solano}, {Solitro}, {Souami}, {Souchay}, {Spagna}, {Spoto}, {Steele},
  {Steidelm{\"u}ller}, {Stephenson}, {S{\"u}veges}, {Szabados}, {Szegedi-Elek},
  {Taris}, {Tauran}, {Taylor}, {Teixeira}, {Thuillot}, {Tonello}, {Torra},
  {Torra}, {Turon}, {Unger}, {Vaillant}, {van Dillen}, {Vanel}, {Vecchiato},
  {Viala}, {Vicente}, {Voutsinas}, {Weiler}, {Wevers}, {Wyrzykowski}, {Yoldas},
  {Yvard}, {Zhao}, {Zorec}, {Zucker}, {Zurbach}, \& {Zwitter}}]{Gaia21}
{Gaia Collaboration}, {Brown}, A.~G.~A., {Vallenari}, A., {et~al.} 2021, \aap,
  649, A1

\bibitem[{{Gaia Collaboration} {et~al.}(2018){Gaia Collaboration}, {Helmi},
  {van Leeuwen}, {McMillan}, {Massari}, {Antoja}, {Robin}, {Lindegren},
  {Bastian}, {Arenou}, {Babusiaux}, {Biermann}, {Breddels}, {Hobbs}, {Jordi},
  {Pancino}, {Reyl{\'e}}, {Veljanoski}, {Brown}, {Vallenari}, {Prusti}, {de
  Bruijne}, {Bailer-Jones}, {Evans}, {Eyer}, {Jansen}, {Klioner}, {Lammers},
  {Luri}, {Mignard}, {Panem}, {Pourbaix}, {Randich}, {Sartoretti}, {Siddiqui},
  {Soubiran}, {Walton}, {Cropper}, {Drimmel}, {Katz}, {Lattanzi}, {Bakker},
  {Cacciari}, {Casta{\~n}eda}, {Chaoul}, {Cheek}, {De Angeli}, {Fabricius},
  {Guerra}, {Holl}, {Masana}, {Messineo}, {Mowlavi}, {Nienartowicz}, {Panuzzo},
  {Portell}, {Riello}, {Seabroke}, {Tanga}, {Th{\'e}venin}, {Gracia-Abril},
  {Comoretto}, {Garcia-Reinaldos}, {Teyssier}, {Altmann}, {Andrae}, {Audard},
  {Bellas-Velidis}, {Benson}, {Berthier}, {Blomme}, {Burgess}, {Busso},
  {Carry}, {Cellino}, {Clementini}, {Clotet}, {Creevey}, {Davidson}, {De
  Ridder}, {Delchambre}, {Dell'Oro}, {Ducourant},
  {Fern{\'a}ndez-Hern{\'a}ndez}, {Fouesneau}, {Fr{\'e}mat}, {Galluccio},
  {Garc{\'\i}a-Torres}, {Gonz{\'a}lez-N{\'u}{\~n}ez}, {Gonz{\'a}lez-Vidal},
  {Gosset}, {Guy}, {Halbwachs}, {Hambly}, {Harrison}, {Hern{\'a}ndez},
  {Hestroffer}, {Hodgkin}, {Hutton}, {Jasniewicz}, {Jean-Antoine-Piccolo},
  {Jordan}, {Korn}, {Krone-Martins}, {Lanzafame}, {Lebzelter}, {L{\"o}ffler},
  {Manteiga}, {Marrese}, {Mart{\'\i}n-Fleitas}, {Moitinho}, {Mora}, {Muinonen},
  {Osinde}, {Pauwels}, {Petit}, {Recio-Blanco}, {Richards}, {Rimoldini},
  {Sarro}, {Siopis}, {Smith}, {Sozzetti}, {S{\"u}veges}, {Torra}, {van Reeven},
  {Abbas}, {Abreu Aramburu}, {Accart}, {Aerts}, {Altavilla}, {{\'A}lvarez},
  {Alvarez}, {Alves}, {Anderson}, {Andrei}, {Anglada Varela}, {Antiche},
  {Arcay}, {Astraatmadja}, {Bach}, {Baker}, {Balaguer-N{\'u}{\~n}ez}, {Balm},
  {Barache}, {Barata}, {Barbato}, {Barblan}, {Barklem}, {Barrado}, {Barros},
  {Barstow}, {Bartholom{\'e} Mu{\~n}oz}, {Bassilana}, {Becciani}, {Bellazzini},
  {Berihuete}, {Bertone}, {Bianchi}, {Bienaym{\'e}}, {Blanco-Cuaresma}, {Boch},
  {Boeche}, {Bombrun}, {Borrachero}, {Bossini}, {Bouquillon}, {Bourda},
  {Bragaglia}, {Bramante}, {Bressan}, {Brouillet}, {Br{\"u}semeister},
  {Brugaletta}, {Bucciarelli}, {Burlacu}, {Busonero}, {Butkevich}, {Buzzi},
  {Caffau}, {Cancelliere}, {Cannizzaro}, {Cantat-Gaudin}, {Carballo},
  {Carlucci}, {Carrasco}, {Casamiquela}, {Castellani}, {Castro-Ginard},
  {Charlot}, {Chemin}, {Chiavassa}, {Cocozza}, {Costigan}, {Cowell}, {Crifo},
  {Crosta}, {Crowley}, {Cuypers}, {Dafonte}, {Damerdji}, {Dapergolas}, {David},
  {David}, {de Laverny}, {De Luise}, {De March}, {de Martino}, {de Souza}, {de
  Torres}, {Debosscher}, {del Pozo}, {Delbo}, {Delgado}, {Delgado}, {Di
  Matteo}, {Diakite}, {Diener}, {Distefano}, {Dolding}, {Drazinos},
  {Dur{\'a}n}, {Edvardsson}, {Enke}, {Eriksson}, {Esquej}, {Eynard Bontemps},
  {Fabre}, {Fabrizio}, {Faigler}, {Falc{\~a}o}, {Farr{\`a}s Casas}, {Federici},
  {Fedorets}, {Fernique}, {Figueras}, {Filippi}, {Findeisen}, {Fonti},
  {Fraile}, {Fraser}, {Fr{\'e}zouls}, {Gai}, {Galleti}, {Garabato},
  {Garc{\'\i}a-Sedano}, {Garofalo}, {Garralda}, {Gavel}, {Gavras}, {Gerssen},
  {Geyer}, {Giacobbe}, {Gilmore}, {Girona}, {Giuffrida}, {Glass}, {Gomes},
  {Granvik}, {Gueguen}, {Guerrier}, {Guiraud}, {Guti{\'e}rrez-S{\'a}nchez},
  {Hofmann}, {Holland}, {Huckle}, {Hypki}, {Icardi}, {Jan{\ss}en}, {Jevardat de
  Fombelle}, {Jonker}, {Juh{\'a}sz}, {Julbe}, {Karampelas}, {Kewley}, {Klar},
  {Kochoska}, {Kohley}, {Kolenberg}, {Kontizas}, {Kontizas}, {Koposov},
  {Kordopatis}, {Kostrzewa-Rutkowska}, {Koubsky}, {Lambert}, {Lanza}, {Lasne},
  {Lavigne}, {Le Fustec}, {Le Poncin-Lafitte}, {Lebreton}, {Leccia}, {Leclerc},
  {Lecoeur-Taibi}, {Lenhardt}, {Leroux}, {Liao}, {Licata}, {Lindstr{\o}m},
  {Lister}, {Livanou}, {Lobel}, {L{\'o}pez}, {Managau}, {Mann}, {Mantelet},
  {Marchal}, {Marchant}, {Marconi}, {Marinoni}, {Marschalk{\'o}}, {Marshall},
  {Martino}, {Marton}, {Mary}, {Matijevi{\v{c}}}, {Mazeh}, {Messina},
  {Michalik}, {Millar}, {Molina}, {Molinaro}, {Moln{\'a}r}, {Montegriffo},
  {Mor}, {Morbidelli}, {Morel}, {Morris}, {Mulone}, {Muraveva}, {Musella},
  {Nelemans}, {Nicastro}, {Noval}, {O'Mullane}, {Ord{\'e}novic},
  {Ord{\'o}{\~n}ez-Blanco}, {Osborne}, {Pagani}, {Pagano}, {Pailler},
  {Palacin}, {Palaversa}, {Panahi}, {Pawlak}, {Piersimoni}, {Pineau}, {Plachy},
  {Plum}, {Poggio}, {Poujoulet}, {Pr{\v{s}}a}, {Pulone}, {Racero}, {Ragaini},
  {Rambaux}, {Ramos-Lerate}, {Regibo}, {Riclet}, {Ripepi}, {Riva}, {Rivard},
  {Rixon}, {Roegiers}, {Roelens}, {Romero-G{\'o}mez}, {Rowell}, {Royer},
  {Ruiz-Dern}, {Sadowski}, {Sagrist{\`a} Sell{\'e}s}, {Sahlmann}, {Salgado},
  {Salguero}, {Sanna}, {Santana-Ros}, {Sarasso}, {Savietto}, {Schultheis},
  {Sciacca}, {Segol}, {Segovia}, {S{\'e}gransan}, {Shih}, {Siltala}, {Silva},
  {Smart}, {Smith}, {Solano}, {Solitro}, {Sordo}, {Soria Nieto}, {Souchay},
  {Spagna}, {Spoto}, {Stampa}, {Steele}, {Steidelm{\"u}ller}, {Stephenson},
  {Stoev}, {Suess}, {Surdej}, {Szabados}, {Szegedi-Elek}, {Tapiador}, {Taris},
  {Tauran}, {Taylor}, {Teixeira}, {Terrett}, {Teyssandier}, {Thuillot},
  {Titarenko}, {Torra Clotet}, {Turon}, {Ulla}, {Utrilla}, {Uzzi}, {Vaillant},
  {Valentini}, {Valette}, {van Elteren}, {Van Hemelryck}, {van Leeuwen},
  {Vaschetto}, {Vecchiato}, {Viala}, {Vicente}, {Vogt}, {von Essen}, {Voss},
  {Votruba}, {Voutsinas}, {Walmsley}, {Weiler}, {Wertz}, {Wevems},
  {Wyrzykowski}, {Yoldas}, {{\v{Z}}erjal}, {Ziaeepour}, {Zorec}, {Zschocke},
  {Zucker}, {Zurbach}, \& {Zwitter}}]{Gaia18b}
{Gaia Collaboration}, {Helmi}, A., {van Leeuwen}, F., {et~al.} 2018, \aap, 616,
  A12

\bibitem[{{Garcia} {et~al.}(2021){Garcia}, {Evans}, {Bestenlehner}, {Bouret},
  {Castro}, {Cervi{\~n}o}, {Fullerton}, {Gieles}, {Herrero}, {de Koter},
  {Lennon}, {van Loon}, {Martins}, {de Mink}, {Najarro}, {Negueruela}, {Sana},
  {Sim{\'o}n-D{\'\i}az}, {Sz{\'e}csi}, {Tramper}, {Vink}, \&
  {Wofford}}]{Garcia21}
{Garcia}, M., {Evans}, C.~J., {Bestenlehner}, J.~M., {et~al.} 2021,
  Experimental Astronomy, 51, 887

\bibitem[{{Garcia} {et~al.}(2017){Garcia}, {Herrero}, {Najarro}, {Camacho},
  {Lennon}, {Urbaneja}, \& {Castro}}]{Garcia17}
{Garcia}, M., {Herrero}, A., {Najarro}, F., {et~al.} 2017, in The Lives and
  Death-Throes of Massive Stars, ed. J.~J. {Eldridge}, J.~C. {Bray}, L.~A.~S.
  {McClelland}, \& L.~{Xiao}, Vol. 329, 313--321

\bibitem[{{Gehan} {et~al.}(2021){Gehan}, {Mosser}, {Michel}, \&
  {Cunha}}]{Gehan21}
{Gehan}, C., {Mosser}, B., {Michel}, E., \& {Cunha}, M.~S. 2021, \aap, 645,
  A124

\bibitem[{{Gordon} {et~al.}(2003){Gordon}, {Clayton}, {Misselt}, {Landolt}, \&
  {Wolff}}]{Gordon03}
{Gordon}, K.~D., {Clayton}, G.~C., {Misselt}, K.~A., {Landolt}, A.~U., \&
  {Wolff}, M.~J. 2003, \apj, 594, 279

\bibitem[{{Hainich} {et~al.}(2015){Hainich}, {Pasemann}, {Todt}, {Shenar},
  {Sander}, \& {Hamann}}]{Hainich15}
{Hainich}, R., {Pasemann}, D., {Todt}, H., {et~al.} 2015, \aap, 581, A21

\bibitem[{{Hastings} {et~al.}(2021){Hastings}, {Langer}, {Wang},
  {Schootemeijer}, \& {Milone}}]{Hastings21}
{Hastings}, B., {Langer}, N., {Wang}, C., {Schootemeijer}, A., \& {Milone},
  A.~P. 2021, arXiv e-prints, arXiv:2106.12263

\bibitem[{{Hastings} {et~al.}(2020){Hastings}, {Wang}, \&
  {Langer}}]{Hastings20}
{Hastings}, B., {Wang}, C., \& {Langer}, N. 2020, \aap, 633, A165

\bibitem[{{Hilditch} {et~al.}(2005){Hilditch}, {Howarth}, \&
  {Harries}}]{Hilditch05}
{Hilditch}, R.~W., {Howarth}, I.~D., \& {Harries}, T.~J. 2005, \mnras, 357, 304

\bibitem[{{Holtzman} {et~al.}(2006){Holtzman}, {Afonso}, \&
  {Dolphin}}]{Holtzman06}
{Holtzman}, J.~A., {Afonso}, C., \& {Dolphin}, A. 2006, \apjs, 166, 534

\bibitem[{{Hummel} {et~al.}(1999){Hummel}, {Szeifert}, {G{\"a}ssler},
  {Muschielok}, {Seifert}, {Appenzeller}, \& {Rupprecht}}]{Hummel99}
{Hummel}, W., {Szeifert}, T., {G{\"a}ssler}, W., {et~al.} 1999, \aap, 352, L31

\bibitem[{{Hunter} {et~al.}(2008){Hunter}, {Lennon}, {Dufton}, {Trundle},
  {Sim{\'o}n-D{\'{\i}}az}, {Smartt}, {Ryans}, \& {Evans}}]{Hunter08b}
{Hunter}, I., {Lennon}, D.~J., {Dufton}, P.~L., {et~al.} 2008, \aap, 479, 541

\bibitem[{{Iqbal} \& {Keller}(2013)}]{Iqbal13}
{Iqbal}, S. \& {Keller}, S.~C. 2013, \mnras, 435, 3103

\bibitem[{{Jackson} \& {Jeffries}(2010)}]{Jackson10}
{Jackson}, R.~J. \& {Jeffries}, R.~D. 2010, \mnras, 402, 1380

\bibitem[{{Kaaret} {et~al.}(2017){Kaaret}, {Feng}, \& {Roberts}}]{Kaaret17}
{Kaaret}, P., {Feng}, H., \& {Roberts}, T.~P. 2017, \araa, 55, 303

\bibitem[{{Karachentsev} {et~al.}(2002){Karachentsev}, {Dolphin}, {Geisler},
  {Grebel}, {Guhathakurta}, {Hodge}, {Karachentseva}, {Sarajedini}, {Seitzer},
  \& {Sharina}}]{Karachentsev02}
{Karachentsev}, I.~D., {Dolphin}, A.~E., {Geisler}, D., {et~al.} 2002, \aap,
  383, 125

\bibitem[{{Kaufer} {et~al.}(2004){Kaufer}, {Venn}, {Tolstoy}, {Pinte}, \&
  {Kudritzki}}]{Kaufer04}
{Kaufer}, A., {Venn}, K.~A., {Tolstoy}, E., {Pinte}, C., \& {Kudritzki}, R.-P.
  2004, \aj, 127, 2723

\bibitem[{{Keller} {et~al.}(1999){Keller}, {Wood}, \& {Bessell}}]{Keller99}
{Keller}, S.~C., {Wood}, P.~R., \& {Bessell}, M.~S. 1999, \aaps, 134, 489

\bibitem[{{Klement} {et~al.}(2019){Klement}, {Carciofi}, {Rivinius}, {Ignace},
  {Matthews}, {Torstensson}, {Gies}, {Vieira}, {Richardson}, {Domiciano de
  Souza}, {Bjorkman}, {Hallinan}, {Faes}, {Mota}, {Gullingsrud}, {de Breuck},
  {Kervella}, {Cur{\'e}}, \& {Gunawan}}]{Klement19}
{Klement}, R., {Carciofi}, A.~C., {Rivinius}, T., {et~al.} 2019, \apj, 885, 147

\bibitem[{{Kniazev} {et~al.}(2005){Kniazev}, {Grebel}, {Pustilnik}, {Pramskij},
  \& {Zucker}}]{Kniazev05}
{Kniazev}, A.~Y., {Grebel}, E.~K., {Pustilnik}, S.~A., {Pramskij}, A.~G., \&
  {Zucker}, D.~B. 2005, \aj, 130, 1558

\bibitem[{{Korn} {et~al.}(2000){Korn}, {Becker}, {Gummersbach}, \&
  {Wolf}}]{Korn00}
{Korn}, A.~J., {Becker}, S.~R., {Gummersbach}, C.~A., \& {Wolf}, B. 2000, \aap,
  353, 655

\bibitem[{{Kruckow} {et~al.}(2018){Kruckow}, {Tauris}, {Langer}, {Kramer}, \&
  {Izzard}}]{Kruckow18}
{Kruckow}, M.~U., {Tauris}, T.~M., {Langer}, N., {Kramer}, M., \& {Izzard},
  R.~G. 2018, \mnras, 481, 1908

\bibitem[{{Langer}(1992)}]{Langer92}
{Langer}, N. 1992, \aap, 265, L17

\bibitem[{{Lunnan} {et~al.}(2014){Lunnan}, {Chornock}, {Berger}, {Laskar},
  {Fong}, {Rest}, {Sanders}, {Challis}, {Drout}, {Foley}, {Huber}, {Kirshner},
  {Leibler}, {Marion}, {McCrum}, {Milisavljevic}, {Narayan}, {Scolnic},
  {Smartt}, {Smith}, {Soderberg}, {Tonry}, {Burgett}, {Chambers}, {Flewelling},
  {Hodapp}, {Kaiser}, {Magnier}, {Price}, \& {Wainscoat}}]{Lunnan14}
{Lunnan}, R., {Chornock}, R., {Berger}, E., {et~al.} 2014, \apj, 787, 138

\bibitem[{{Macri} {et~al.}(2006){Macri}, {Stanek}, {Bersier}, {Greenhill}, \&
  {Reid}}]{Macri06}
{Macri}, L.~M., {Stanek}, K.~Z., {Bersier}, D., {Greenhill}, L.~J., \& {Reid},
  M.~J. 2006, \apj, 652, 1133

\bibitem[{{Maeder}(1987)}]{Maeder87}
{Maeder}, A. 1987, \aap, 178, 159

\bibitem[{{Maeder} {et~al.}(1999){Maeder}, {Grebel}, \&
  {Mermilliod}}]{Maeder99}
{Maeder}, A., {Grebel}, E.~K., \& {Mermilliod}, J.-C. 1999, \aap, 346, 459

\bibitem[{{Marchant} {et~al.}(2017){Marchant}, {Langer}, {Podsiadlowski},
  {Tauris}, {de Mink}, {Mandel}, \& {Moriya}}]{Marchant17}
{Marchant}, P., {Langer}, N., {Podsiadlowski}, P., {et~al.} 2017, \aap, 604,
  A55

\bibitem[{{Marchant} {et~al.}(2016){Marchant}, {Langer}, {Podsiadlowski},
  {Tauris}, \& {Moriya}}]{Marchant16}
{Marchant}, P., {Langer}, N., {Podsiadlowski}, P., {Tauris}, T.~M., \&
  {Moriya}, T.~J. 2016, \aap, 588, A50

\bibitem[{{Martayan} {et~al.}(2010){Martayan}, {Baade}, \&
  {Fabregat}}]{Martayan10}
{Martayan}, C., {Baade}, D., \& {Fabregat}, J. 2010, \aap, 509, A11

\bibitem[{{Martayan} {et~al.}(2006){Martayan}, {Fr{\'e}mat}, {Hubert},
  {Floquet}, {Zorec}, \& {Neiner}}]{Martayan06}
{Martayan}, C., {Fr{\'e}mat}, Y., {Hubert}, A.~M., {et~al.} 2006, \aap, 452,
  273

\bibitem[{{Martayan} {et~al.}(2007){Martayan}, {Fr{\'e}mat}, {Hubert},
  {Floquet}, {Zorec}, \& {Neiner}}]{Martayan07}
{Martayan}, C., {Fr{\'e}mat}, Y., {Hubert}, A.~M., {et~al.} 2007, \aap, 462,
  683

\bibitem[{{Mateo}(1998)}]{Mateo98}
{Mateo}, M.~L. 1998, \araa, 36, 435

\bibitem[{{McSwain} \& {Gies}(2005)}]{McSwain05}
{McSwain}, M.~V. \& {Gies}, D.~R. 2005, \apjs, 161, 118

\bibitem[{{Meixner} {et~al.}(2006){Meixner}, {Gordon}, {Indebetouw}, {Hora},
  {Whitney}, {Blum}, {Reach}, {Bernard}, {Meade}, {Babler}, {Engelbracht},
  {For}, {Misselt}, {Vijh}, {Leitherer}, {Cohen}, {Churchwell}, {Boulanger},
  {Frogel}, {Fukui}, {Gallagher}, {Gorjian}, {Harris}, {Kelly}, {Kawamura},
  {Kim}, {Latter}, {Madden}, {Markwick-Kemper}, {Mizuno}, {Mizuno}, {Mould},
  {Nota}, {Oey}, {Olsen}, {Onishi}, {Paladini}, {Panagia}, {Perez-Gonzalez},
  {Shibai}, {Sato}, {Smith}, {Staveley-Smith}, {Tielens}, {Ueta}, {van Dyk},
  {Volk}, {Werner}, \& {Zaritsky}}]{Meixner06}
{Meixner}, M., {Gordon}, K.~D., {Indebetouw}, R., {et~al.} 2006, \aj, 132, 2268

\bibitem[{{Milone} {et~al.}(2018){Milone}, {Marino}, {Di Criscienzo},
  {D'Antona}, {Bedin}, {Da Costa}, {Piotto}, {Tailo}, {Dotter}, {Angeloni},
  {Anderson}, {Jerjen}, {Li}, {Dupree}, {Granata}, {Lagioia}, {Mackey},
  {Nardiello}, \& {Vesperini}}]{Milone18}
{Milone}, A.~P., {Marino}, A.~F., {Di Criscienzo}, M., {et~al.} 2018, \mnras,
  477, 2640

\bibitem[{{Mokiem} {et~al.}(2006){Mokiem}, {de Koter}, {Evans}, {Puls},
  {Smartt}, {Crowther}, {Herrero}, {Langer}, {Lennon}, {Najarro}, {Villamariz},
  \& {Yoon}}]{Mokiem06}
{Mokiem}, M.~R., {de Koter}, A., {Evans}, C.~J., {et~al.} 2006, \aap, 456, 1131

\bibitem[{{Mokiem} {et~al.}(2007){Mokiem}, {de Koter}, {Vink}, {Puls}, {Evans},
  {Smartt}, {Crowther}, {Herrero}, {Langer}, {Lennon}, {Najarro}, \&
  {Villamariz}}]{Mokiem07}
{Mokiem}, M.~R., {de Koter}, A., {Vink}, J.~S., {et~al.} 2007, \aap, 473, 603

\bibitem[{{Mosser} {et~al.}(2018){Mosser}, {Gehan}, {Belkacem}, {Samadi},
  {Michel}, \& {Goupil}}]{Mosser18}
{Mosser}, B., {Gehan}, C., {Belkacem}, K., {et~al.} 2018, \aap, 618, A109

\bibitem[{{Neugent} {et~al.}(2018){Neugent}, {Massey}, \&
  {Morrell}}]{Neugent18}
{Neugent}, K.~F., {Massey}, P., \& {Morrell}, N. 2018, \apj, 863, 181

\bibitem[{{Okazaki}(2001)}]{Okazaki01}
{Okazaki}, A.~T. 2001, \pasj, 53, 119

\bibitem[{{Peters} {et~al.}(2020){Peters}, {Wisniewski}, {Williams}, {Lomax},
  {Choi}, {Durbin}, {Johnson}, {Lewis}, {Lutz}, {Sigut}, {Wallach}, \&
  {Dalcanton}}]{Peters20}
{Peters}, M., {Wisniewski}, J.~P., {Williams}, B.~F., {et~al.} 2020, \aj, 159,
  119

\bibitem[{{Pineau} {et~al.}(2020){Pineau}, {Boch}, {Derri{\`e}re}, \&
  {Schaaff}}]{Pineau20}
{Pineau}, F.-X., {Boch}, T., {Derri{\`e}re}, S., \& {Schaaff}, A. 2020, in
  Astronomical Society of the Pacific Conference Series, Vol. 522, Astronomical
  Data Analysis Software and Systems XXVII, ed. P.~{Ballester}, J.~{Ibsen},
  M.~{Solar}, \& K.~{Shortridge}, 125

\bibitem[{{Ramachandran} {et~al.}(2018){Ramachandran}, {Hamann}, {Hainich},
  {Oskinova}, {Shenar}, {Sander}, {Todt}, \& {Gallagher}}]{Ramachandran18}
{Ramachandran}, V., {Hamann}, W.~R., {Hainich}, R., {et~al.} 2018, \aap, 615,
  A40

\bibitem[{{Ramachandran} {et~al.}(2019){Ramachandran}, {Hamann}, {Oskinova},
  {Gallagher}, {Hainich}, {Shenar}, {Sand er}, {Todt}, \&
  {Fulmer}}]{Ramachandran19}
{Ramachandran}, V., {Hamann}, W.~R., {Oskinova}, L.~M., {et~al.} 2019, \aap,
  625, A104

\bibitem[{{Rivinius} {et~al.}(2013){Rivinius}, {Carciofi}, \&
  {Martayan}}]{Rivinius13}
{Rivinius}, T., {Carciofi}, A.~C., \& {Martayan}, C. 2013, \aapr, 21, 69

\bibitem[{{Sabbi} {et~al.}(2020){Sabbi}, {Adamo}, {Bianchi}, {Calzetti},
  {Cignoni}, {Crowther}, {Eldridge}, {Elmegreen}, {Elmegreen}, {Gallagher},
  {Garcia}, {Gouliermis}, {Grebel}, {Klessen}, {Nota}, {Smith}, {Wofford}, \&
  {Zeidler}}]{Sabbi20}
{Sabbi}, E., {Adamo}, A., {Bianchi}, L.~C., {et~al.} 2020, {GULP: Galaxy UV
  Legacy Project}, HST Proposal. Cycle 28, ID. \#16316

\bibitem[{{Sabbi} {et~al.}(2018){Sabbi}, {Calzetti}, {Ubeda}, {Adamo},
  {Cignoni}, {Thilker}, {Aloisi}, {Elmegreen}, {Elmegreen}, {Gouliermis},
  {Grebel}, {Messa}, {Smith}, {Tosi}, {Dolphin}, {Andrews}, {Ashworth},
  {Bright}, {Brown}, {Chandar}, {Christian}, {Clayton}, {Cook}, {Dale}, {de
  Mink}, {Dobbs}, {Evans}, {Fumagalli}, {Gallagher}, {Grasha}, {Herrero},
  {Hunter}, {Johnson}, {Kahre}, {Kennicutt}, {Kim}, {Krumholz}, {Lee},
  {Lennon}, {Martin}, {Nair}, {Nota}, {{\"O}stlin}, {Pellerin}, {Prieto},
  {Regan}, {Ryon}, {Sacchi}, {Schaerer}, {Schiminovich}, {Shabani}, {Van Dyk},
  {Walterbos}, {Whitmore}, \& {Wofford}}]{Sabbi18}
{Sabbi}, E., {Calzetti}, D., {Ubeda}, L., {et~al.} 2018, \apjs, 235, 23

\bibitem[{{Sana} {et~al.}(2012){Sana}, {de Mink}, {de Koter}, {Langer},
  {Evans}, {Gieles}, {Gosset}, {Izzard}, {Le Bouquin}, \& {Schneider}}]{Sana12}
{Sana}, H., {de Mink}, S.~E., {de Koter}, A., {et~al.} 2012, Science, 337, 444

\bibitem[{{Savaglio} {et~al.}(2009){Savaglio}, {Glazebrook}, \& {Le
  Borgne}}]{Savaglio09}
{Savaglio}, S., {Glazebrook}, K., \& {Le Borgne}, D. 2009, \apj, 691, 182

\bibitem[{{Saviane} {et~al.}(2002){Saviane}, {Rizzi}, {Held}, {Bresolin}, \&
  {Momany}}]{Saviane02}
{Saviane}, I., {Rizzi}, L., {Held}, E.~V., {Bresolin}, F., \& {Momany}, Y.
  2002, \aap, 390, 59

\bibitem[{{Schneider} {et~al.}(2019){Schneider}, {Ohlmann}, {Podsiadlowski},
  {R{\"o}pke}, {Balbus}, {Pakmor}, \& {Springel}}]{Schneider19}
{Schneider}, F. R.~N., {Ohlmann}, S.~T., {Podsiadlowski}, P., {et~al.} 2019,
  \nat, 574, 211

\bibitem[{{Schootemeijer} {et~al.}(2018){Schootemeijer}, {G{\"o}tberg}, {Mink},
  {Gies}, \& {Zapartas}}]{Schootemeijer18b}
{Schootemeijer}, A., {G{\"o}tberg}, Y., {Mink}, S.~E.~d., {Gies}, D., \&
  {Zapartas}, E. 2018, \aap, 615, A30

\bibitem[{{Schootemeijer} {et~al.}(2019){Schootemeijer}, {Langer}, {Grin}, \&
  {Wang}}]{Schootemeijer19}
{Schootemeijer}, A., {Langer}, N., {Grin}, N.~J., \& {Wang}, C. 2019, \aap,
  625, A132

\bibitem[{{Schootemeijer} {et~al.}(2021){Schootemeijer}, {Langer}, {Lennon},
  {Evans}, {Crowther}, {Geen}, {Howarth}, {de Koter}, {Menten}, \&
  {Vink}}]{Schootemeijer21}
{Schootemeijer}, A., {Langer}, N., {Lennon}, D., {et~al.} 2021, \aap, 646, A106

\bibitem[{{Shenar} {et~al.}(2020){Shenar}, {Gilkis}, {Vink}, {Sana}, \&
  {Sander}}]{Shenar20}
{Shenar}, T., {Gilkis}, A., {Vink}, J.~S., {Sana}, H., \& {Sander}, A.~A.~C.
  2020, \aap, 634, A79

\bibitem[{{Struve}(1931)}]{Struve31}
{Struve}, O. 1931, \apj, 73, 94

\bibitem[{{Sz{\'e}csi} {et~al.}(2015){Sz{\'e}csi}, {Langer}, {Yoon}, {Sanyal},
  {de Mink}, {Evans}, \& {Dermine}}]{Szecsi15}
{Sz{\'e}csi}, D., {Langer}, N., {Yoon}, S.-C., {et~al.} 2015, \aap, 581, A15

\bibitem[{{Tammann} {et~al.}(2011){Tammann}, {Reindl}, \&
  {Sandage}}]{Tammann11}
{Tammann}, G.~A., {Reindl}, B., \& {Sandage}, A. 2011, \aap, 531, A134

\bibitem[{{Taylor}(2005)}]{Taylor05}
{Taylor}, M.~B. 2005, in Astronomical Society of the Pacific Conference Series,
  Vol. 347, Astronomical Data Analysis Software and Systems XIV, ed.
  P.~{Shopbell}, M.~{Britton}, \& R.~{Ebert}, 29

\bibitem[{{Telford} {et~al.}(2021){Telford}, {Chisholm}, {McQuinn}, \&
  {Berg}}]{Telford21}
{Telford}, O.~G., {Chisholm}, J., {McQuinn}, K. B.~W., \& {Berg}, D.~A. 2021,
  \apj, 922, 191

\bibitem[{{Townsend} {et~al.}(2004){Townsend}, {Owocki}, \&
  {Howarth}}]{Townsend04}
{Townsend}, R.~H.~D., {Owocki}, S.~P., \& {Howarth}, I.~D. 2004, \mnras, 350,
  189

\bibitem[{{Trundle} {et~al.}(2007){Trundle}, {Dufton}, {Hunter}, {Evans},
  {Lennon}, {Smartt}, \& {Ryans}}]{Trundle07}
{Trundle}, C., {Dufton}, P.~L., {Hunter}, I., {et~al.} 2007, \aap, 471, 625

\bibitem[{{van Bever} \& {Vanbeveren}(1997)}]{vanBever97}
{van Bever}, J. \& {Vanbeveren}, D. 1997, \aap, 322, 116

\bibitem[{{van Zee} {et~al.}(2006){van Zee}, {Skillman}, \&
  {Haynes}}]{vanZee06}
{van Zee}, L., {Skillman}, E.~D., \& {Haynes}, M.~P. 2006, \apj, 637, 269

\bibitem[{{Vink} \& {Sander}(2021)}]{Vink21}
{Vink}, J.~S. \& {Sander}, A. A.~C. 2021, \mnras, 504, 2051

\bibitem[{{von Zeipel}(1924)}]{vonZeipel24}
{von Zeipel}, H. 1924, \mnras, 84, 665

\bibitem[{{Wang} {et~al.}(2020){Wang}, {Langer}, {Schootemeijer}, {Castro},
  {Adscheid}, {Marchant}, \& {Hastings}}]{Wang20}
{Wang}, C., {Langer}, N., {Schootemeijer}, A., {et~al.} 2020, \apjl, 888, L12

\bibitem[{{Wisniewski} \& {Bjorkman}(2006)}]{Wisniewski06}
{Wisniewski}, J.~P. \& {Bjorkman}, K.~S. 2006, \apj, 652, 458

\bibitem[{{Yang} {et~al.}(2019){Yang}, {Bonanos}, {Jiang}, {Gao}, {Gavras},
  {Maravelias}, {Ren}, {Wang}, {Xue}, {Tramper}, {Spetsieri}, \&
  {Pouliasis}}]{Yang19}
{Yang}, M., {Bonanos}, A.~Z., {Jiang}, B.-W., {et~al.} 2019, arXiv e-prints,
  arXiv:1907.06717

\bibitem[{{Zaritsky} {et~al.}(2004){Zaritsky}, {Harris}, {Thompson}, \&
  {Grebel}}]{Zaritsky04}
{Zaritsky}, D., {Harris}, J., {Thompson}, I.~B., \& {Grebel}, E.~K. 2004, \aj,
  128, 1606

\bibitem[{{Zaritsky} {et~al.}(2002){Zaritsky}, {Harris}, {Thompson}, {Grebel},
  \& {Massey}}]{Zaritsky02}
{Zaritsky}, D., {Harris}, J., {Thompson}, I.~B., {Grebel}, E.~K., \& {Massey},
  P. 2002, \aj, 123, 855

\bibitem[{{Zorec} \& {Briot}(1997)}]{Zorec97}
{Zorec}, J. \& {Briot}, D. 1997, \aap, 318, 443

\end{thebibliography}
\bibliographystyle{aa} 


\appendix

\section{Tests \label{sec:appb}}

\subsection{Identification of OBe stars \label{sec:appb_id_obes}}

\begin{figure*}[ht]
\centering
\includegraphics[width=\linewidth]{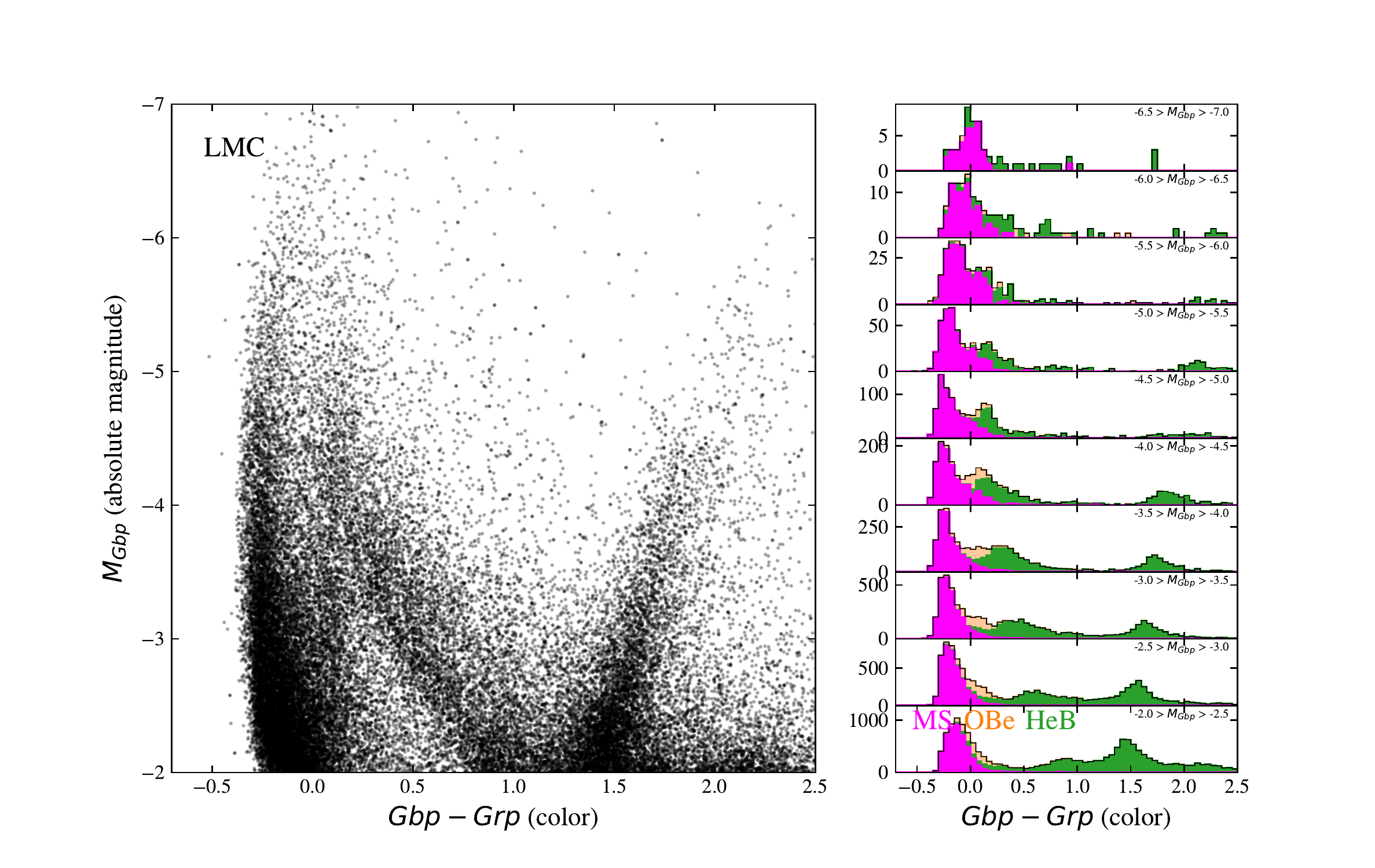}
\caption{Same as Fig.\,\ref{fig:cmd}, but for the Large Magellanic Cloud (LMC). The adopted value for the distance modulus is 18.41 \citep{Macri06}.}
\label{fig:cmd_lmc}
\end{figure*}

\paragraph{Magellanic Clouds} 
To divide the sources in our CMDs of the SMC (Fig.\,\ref{fig:cmd}) and the LMC (Fig.\,\ref{fig:cmd_lmc}) into HeB, MS, and OBe stars, we use color cuts in color-color diagrams (as described in the main text for the SMC). Here we provide a few more details and show how we repeated the method for the LMC.

\begin{figure*}[ht]
\centering
\includegraphics[width=\linewidth]{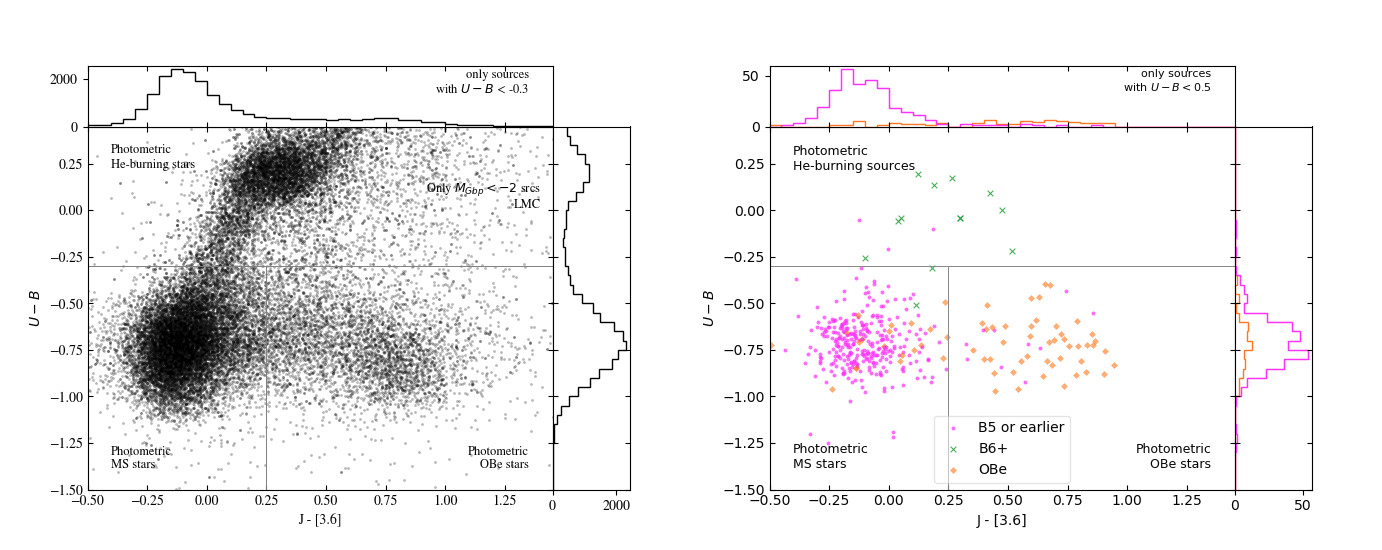}
\caption{Same as Fig.\,\ref{fig:colocolo_3cut}, but now for the Large Magellanic Cloud (LMC). The sources with known spectral types are from \cite{Martayan06, Evans06, Ramachandran18}.
The border between MS and OBe stars lies are the same $J-[3.6]$ color as in the SMC, while the border between He-burning stars and MS/OBe stars lies at $U-B = -0.3$ (for the SMC we adopt $U-B = -0.5$).}
\label{fig:colocolo_3cut_lmc}
\end{figure*}

We base the location of the color cuts in the color-color diagrams (Figs.\,\ref{fig:colocolo_3cut}, \ref{fig:colocolo_3cut_lmc}) on the distribution of stars with known spectral types. The studies that we use are from a sample of spectroscopic SMC studies \citep[][hereafter referred to as SSSS]{Martayan07, Hunter08b, Dufton19, Bonanos10} and a sample of spectroscopic LMC studies \citep[][ hereafter referred to as SSLS]{Evans06, Martayan06, Ramachandran18, Bonanos09}. To be able investigate the colors and magnitudes of the sources presented in these studies, we cross-correlated them with the GAIA EDR3, $UBVI$, and Spitzer-SAGE catalogs described above, again adopting a 1" cross-correlation radius.

From the Bonanos et al. catalogs, hereafter referred to as B09 \citep{Bonanos09} for the LMC catalog and B10 \citep{Bonanos10} for the SMC catalog, in Figs.\ref{fig:colocolo_3cut} and \,\ref{fig:colocolo_3cut_lmc} we show only the stars with spectral type B6 and later (i.e., where we do not make a distinction between stars with and without emission features in their spectrum). We chose this approach because the B09 and B10 catalogs are based on observations that tell apart O/early B and OBe stars less accurately than the rest of the SSLS and SSSS data, as we motivate below. The stars that are classified as OBe stars in B09 and B10 are almost always stars for which H$\alpha$ spectroscopy exists, which is the main diagnostic for the presence of a decretion disk. However, most of the sources in B09 and B10 do not have H$\alpha$ coverage. To illustrate: about 4000 out of 5000 sources in the B10 catalog are from the 2dF survey \citep{Evans04}, and from these 4000 sources about 1000 have been observed in H$\alpha$. A possible OBe nature would most likely not be recognized for the other 3000 sources.

For the location of the color cut between early and late B stars in the SMC we choose $U-B = -0.5$. We do so because above this value most stars with known spectral types are late B stars, while below this value the majority are early B stars (Fig.\,\ref{fig:colocolo_3cut}). We notice a systematic shift in $U-B$ color of about 0.2 magnitudes between the SMC and the LMC (see $U-B$ histograms on the left half of Figs.\,\ref{fig:colocolo_3cut} and \ref{fig:colocolo_3cut_lmc}). Therefore, we choose $U-B = -0.3$ as the color cut between early and late B-type stars in the LMC.

For the color cut between MS and OBe stars in the SMC, we use $J - [3.6] = 0.25$ because at larger values, the number of OBe stars starts to take over the number of non-OBe stars (right panel in Fig.\,\ref{fig:colocolo_3cut}). We notice that unlike for the $U-B$ color, there is no systematic shift in $J - [3.6]$ color distributions between the SMC and the LMC. This could be explained by the small amount of extinction that takes place at these long wavelengths \citep{Gordon03} and the lower sensitivity of the $J - [3.6]$ color (compared to $U-B$) to temperatures of hot stars. Therefore, we adopt $J - [3.6] = 0.25$ as the color cut for OBe stars, as we do for the SMC. 
Also, we note that the resemblance of OBe star $J-[3.6]$ colors in the LMC and SMC could be an indication that the disk properties do not dramatically change between these two environments.

\begin{figure*}[ht]
\centering
\includegraphics[width=0.8\linewidth]{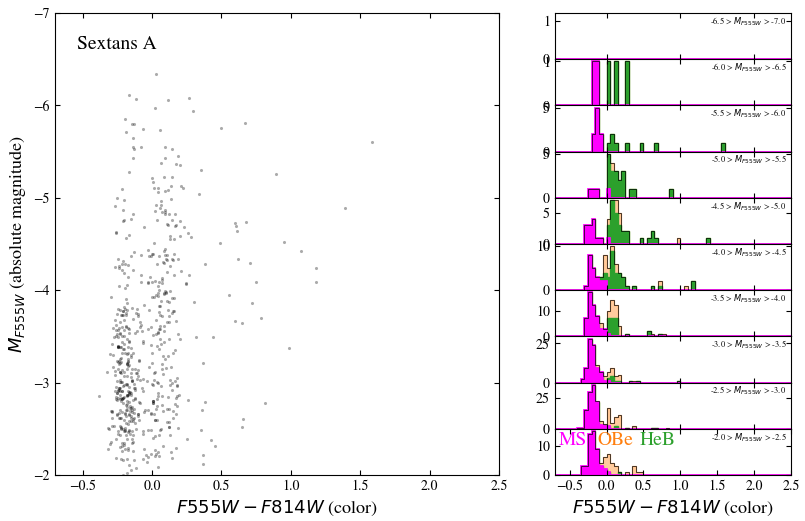}
\caption{Same as Fig.\,\ref{fig:cmd}, but for Hubble Space Telescope data \citep{Holtzman06, Bianchi12} of Sextans A. For this galaxy, 
we identify main-sequence (MS), OBe, and helium-burning (HeB) stars using the color-color diagram Fig.\,\ref{fig:colocolo_sexa}. We adopted a distance modulus of $\mathrm{DM} = 25.63$ \citep{Tammann11}.}
\label{fig:cmd_sexa}
\end{figure*}

\begin{figure*}[ht]
\centering
\includegraphics[width=0.5\linewidth]{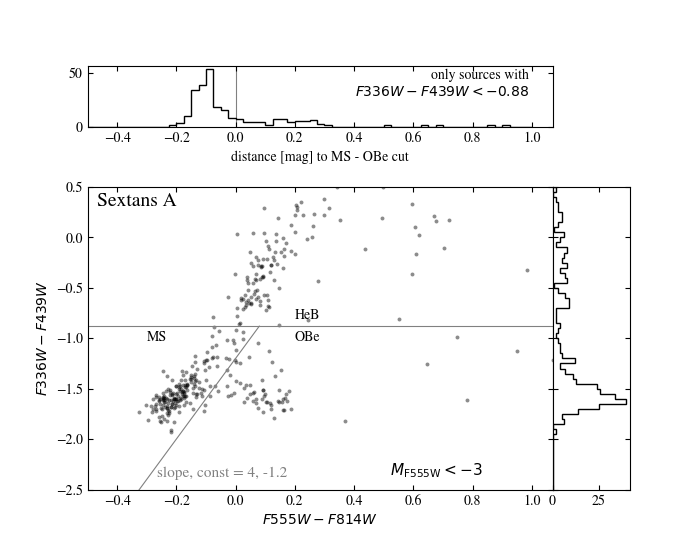}
\caption{Same as the left side of Fig.\,\ref{fig:colocolo_3cut_gaia}, except that the colors are constructed using Hubble Space Telescope data \citep{Holtzman06, Bianchi12} of Sextans A. Only sources brighter than an absolute magnitude of $M_{F555W} = -3$ are shown, for an adopted distance modulus of $\mathrm{DM} = 25.63$ \citep{Tammann11}.}
\label{fig:colocolo_sexa}
\end{figure*}

\paragraph{Sextans A}
The CMD constructed with the Sextans\,A data set is shown in Fig.\,\ref{fig:cmd_sexa} and the color-color diagram in in Fig.\,\ref{fig:colocolo_sexa}. In this diagram a distinct group of stars resides right of the MS stars, as was the case in the SMC and LMC color-color diagrams (Figs.\,\ref{fig:colocolo_3cut} and \ref{fig:colocolo_3cut_lmc}). Since the disk causes a smaller excess for the $F555W - F814W$ color than for the $J-[3.6]$ color, the OBe stars and MS stars are relatively close to each other in Fig.\,\ref{fig:colocolo_sexa}. 
To test our assumption that the three groups in Fig.\,\ref{fig:colocolo_sexa} correspond to MS, OBe, and HeB stars, we provide three different tests. 

\begin{figure*}[ht]
\centering
\includegraphics[width=\linewidth]{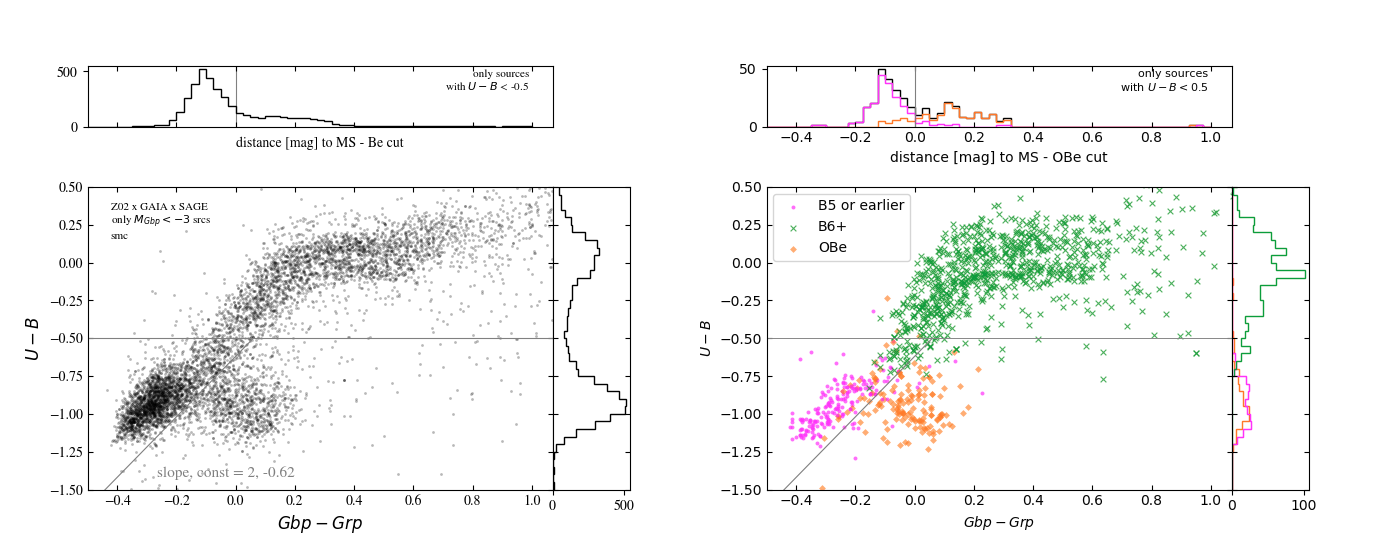}
\caption{\textit{Left:} same as the left panel of Fig.\,\ref{fig:colocolo_3cut}, but on the x-axis is the $Gbp - Grp$ color instead of $J - [3.6]$, and only showing sources brighter than $M_\mathrm{Gbp} = -3$. Also, since we employ a diagonal color cut (grey line), on the x-axis of the top panel we show the distance to the color cut. \textit{Right:} same as the right panel of Fig.\,\ref{fig:colocolo_3cut}, but also showing a $Gbp - Grp$ color instead of $J - [3.6]$.}
\label{fig:colocolo_3cut_gaia}
\end{figure*}

As a first test, in Fig.\,\ref{fig:colocolo_3cut_gaia} we make a similar plot for SSSS data in the SMC (the colors shown on both axes are not exactly the same as in Fig.\,\ref{fig:colocolo_sexa}, but the mean wavelengths of the relevant filters are very similar\footnote{see \url{http://svo2.cab.inta-csic.es/svo/theory/fps/index.php})}. It shows that indeed, in the SMC the MS, OBe and HeB stars are well separated in this diagram, although they are not as segregated as in Fig.\,\ref{fig:colocolo_3cut}. We elect to adopt slightly conservative color cuts for OBe stars in Fig.\,\ref{fig:colocolo_sexa}. In Fig.\,\ref{fig:colocolo_3cut_gaia}, we saw that in the SMC the best location for the color cut lies where the number distribution in the top panels reaches about a quarter of its maximum value $N_\mathrm{max}$. In Fig.\,\ref{fig:colocolo_sexa}, we adopt $N / N_\mathrm{max} = 1/7$.

\begin{figure*}[ht]
\centering
\includegraphics[width=0.7\linewidth]{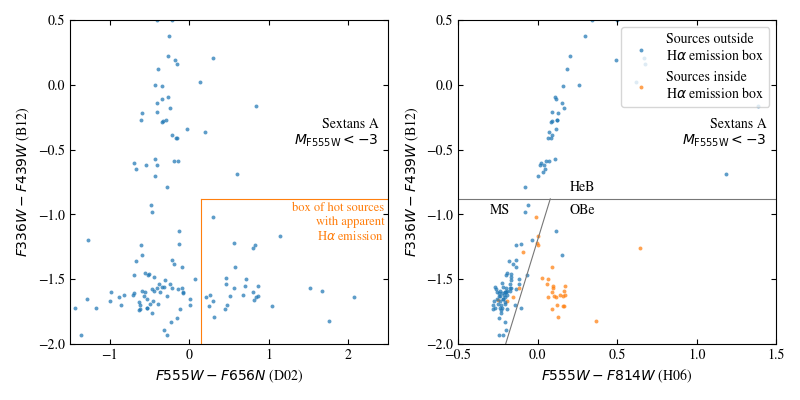}
\caption{
Color-color diagrams of Sextans\,A that are used to test how well we can identify OBe stars there. The catalogs that are used to build photometric colors are provided in parentheses are B12 \citep{Bianchi12}, D02 \citep{Dohm-Palmer02}, and H06 \citep{Holtzman06}.
\textit{Left:} 
Color-color diagram where the color on the x-axis is constructed using the magnitude in the H$\alpha$ narrowband filter $F656N$. Sources that are in the box on the bottom right have colors that indicate that they are hot stars with H$\alpha$ emission (i.e., OBe stars). \textit{Right:} same axes and color cuts as Fig.\,\ref{fig:colocolo_sexa}, but we show sources that are in the box for hot sources with H$\alpha$ emission (left panel) in orange here. The presence of H$\alpha$ excess in most photometric OBe sources supports their inferred OBe nature. 
}
\label{fig:colocolo_sexa_check}
\end{figure*}

As a second test of the effectiveness of the color cuts in Fig.\,\ref{fig:colocolo_sexa}, we investigate a subset of sources in Sextans\,A that have been observed with the H$\alpha$ narrowband filter $F656N$ on board of HST. 
Similar to what we did before, we cross-correlate $F656N$ data from \cite{Dohm-Palmer02} with our B12 x H06 data product in a 0.35" radius. If we cross-match two sources that have a difference larger than 0.4 magnitudes in $F555W$ between the values from \cite{Dohm-Palmer02} and H06, we discard the cross-match.
We present a color-color diagram that includes a color constructed with $F656N$ filter data (Fig.\,\ref{fig:colocolo_sexa_check}).
In the left panel, we draw a box for sources that have a blue $U-B$ color, and a red $F555W - F656N$ color that indicates that they have H$\alpha$ in emission. We expect OBe stars to reside in this box. The right panel of Fig.\,\ref{fig:colocolo_sexa_check} shows that almost all photometric OBe stars appear to have H$\alpha$ in emission, as would be expected.

\begin{figure*}[ht]
\centering
\includegraphics[width=0.4\linewidth]{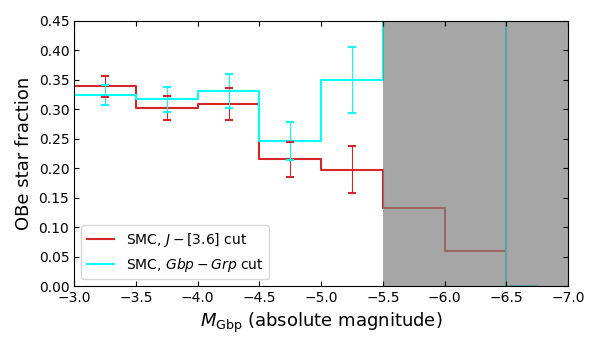}
\caption{OBe star fraction in the SMC as a function of absolute magnitude. The red line shows the result obtained using a color-color cut with the $J - [3.6]$ color on the x-axis (shown in Fig.\,\ref{fig:colocolo_3cut}). As a test, with a cyan line we also show the result obtained when the $Gbp-Grp$ color is used on the x-axis instead (shown in Fig.\,\ref{fig:colocolo_3cut_gaia}).}
\label{fig:fbe_test}
\end{figure*}

One might wonder if the use of an optical/near-IR color for Sextans\,A (Fig.\,\ref{fig:colocolo_sexa}) instead of the $J - [3.6]$ color (e.g., Fig.\,\ref{fig:colocolo_sexa}, for the SMC) to identify OBe stars might systematically affect the derived OBe star fraction. As a third test, we investigate this in the SMC, where both optical/near-IR and $J - [3.6]$ data are available. Figure\,\ref{fig:colocolo_3cut_gaia} is similar to Fig.\,\ref{fig:colocolo_3cut}, except that the color on the x-axis is based on optical/near-IR filters.
We test the impact that the choice of color-color diagram for identifying OBe stars (Fig.\,\ref{fig:colocolo_3cut} versus Fig.\,\ref{fig:colocolo_3cut_gaia}) has on the OBe star fraction that we derive. 
As we show in Fig.\,\ref{fig:fbe_test}, we find that both methods result in highly similar OBe fractions for sources dimmer than $M_\mathrm{Gbp} = -5$. In the brighter-source regime at $M_\mathrm{Gbp} < -5$, the OBe star fraction is higher according to the method that employs the $Gbp - Grp$ color instead of $J - [3.6]$. This could be the result of, for example, the very brightest stars being affected by extinction more than average \citep[see fig.\,6 of][]{Schootemeijer21}, causing them to unjustly being identified as OBe stars. Therefore, we caution that the OBe star fraction in Sextans A in the $-5 > M_\mathrm{abs} > -5.5$ bin (which excluded in Fig.\,\ref{fig:Z_trends}) might in reality be lower than what is shown in Fig.\,\ref{fig:fbe_both}.

\begin{figure*}[ht]
\centering
\includegraphics[width=0.8\linewidth]{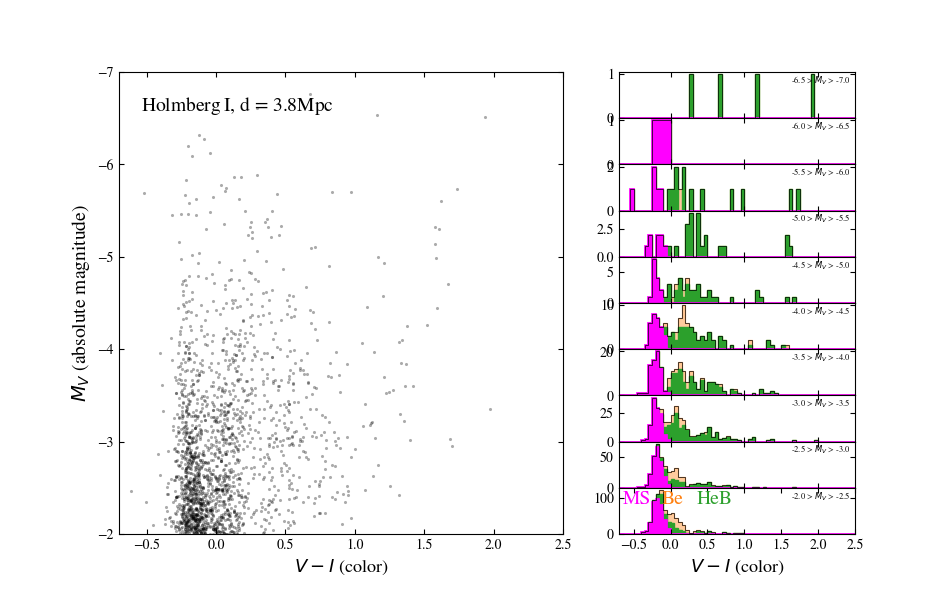}
\caption{Same as Fig.\,\ref{fig:cmd_sexa}, but for Hubble Space Telescope data \citep{Sabbi18} of Holmberg\,I. The adopted distance modulus is 27.92 \citep{Sabbi18}.}
\label{fig:cmd_hoi}
\end{figure*}

\begin{figure*}[ht]
\centering
\includegraphics[width=0.5\linewidth]{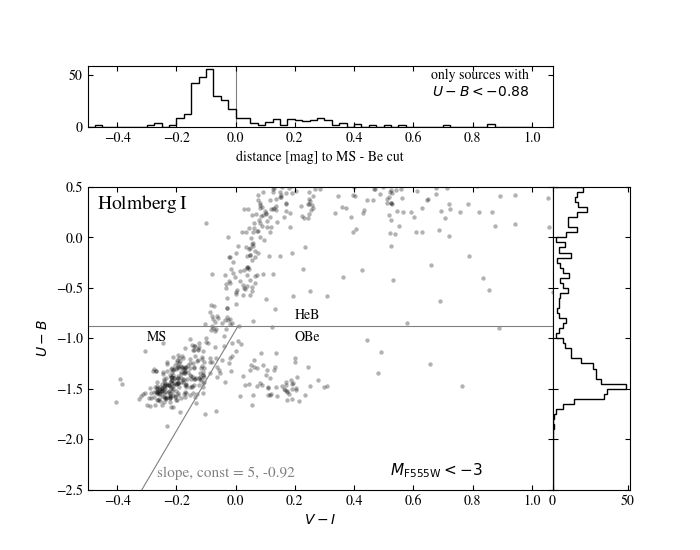}
\caption{ Same as the left panel of Fig.\,\ref{fig:colocolo_3cut_gaia}, except that the colors are constructed using Hubble Space Telescope data \citep{Sabbi18} of Holmberg\,I.}
\label{fig:colocolo_hoi}
\end{figure*}

\begin{figure*}[ht]
\centering
\includegraphics[width=0.8\linewidth]{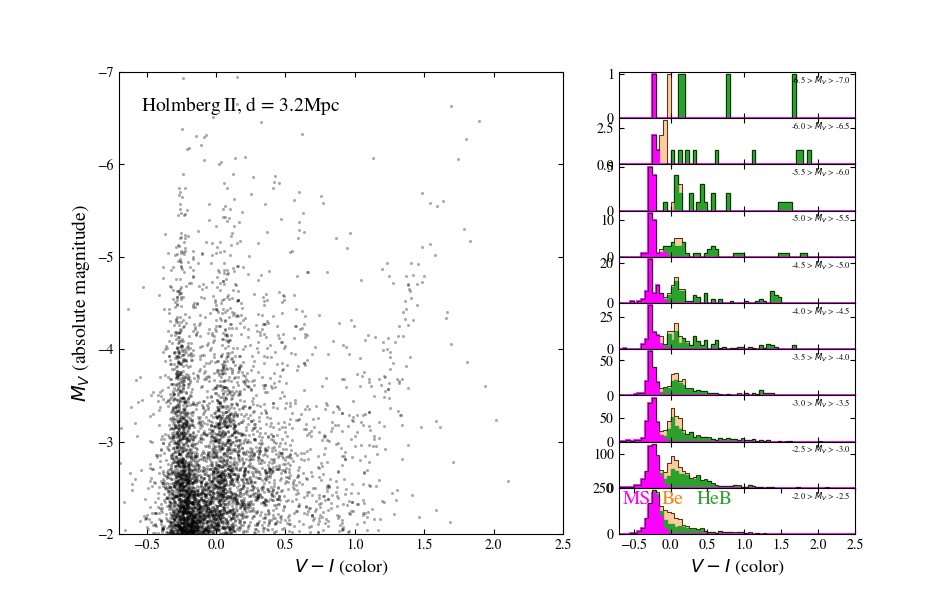}
\caption{Same as Fig.\,\ref{fig:cmd_sexa}, but for Hubble Space Telescope data \citep{Sabbi18} of Holmberg\,II. The adopted distance modulus is 27.55 \citep{Sabbi18}.}
\label{fig:cmd_hoii}
\end{figure*}

\begin{figure*}[ht]
\centering
\includegraphics[width=0.5\linewidth]{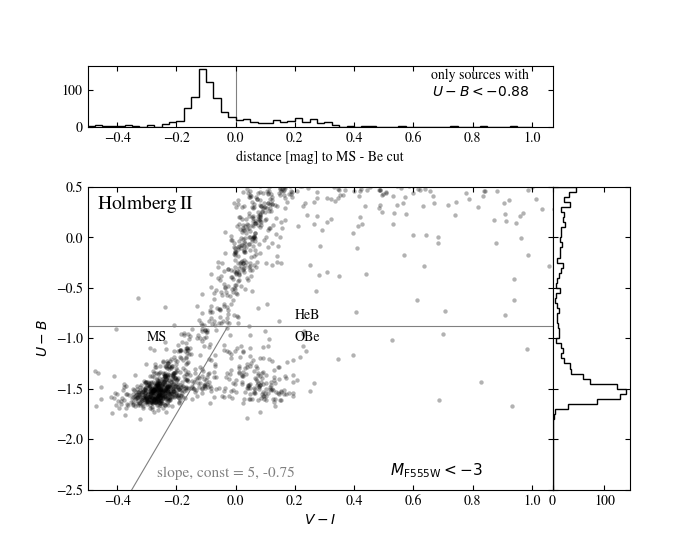}
\caption{  Same as the left side of Fig.\,\ref{fig:colocolo_3cut_gaia}, except that the colors are constructed using Hubble Space Telescope data \citep{Sabbi18} of Holmberg\,II.}
\label{fig:colocolo_hoii}
\end{figure*}

\paragraph{Holmberg\,I and Holmberg\,II}
For Holmberg\,I, we show the CMD in Fig.\,\ref{fig:cmd_hoi} and the color-color diagram in Fig.\,\ref{fig:colocolo_hoi}. For Holmberg\,II, we show the CMD in Fig.\,\ref{fig:cmd_hoii} and the color-color diagram in Fig.\,\ref{fig:colocolo_hoii}. 
In both Fig.\,\ref{fig:colocolo_hoi} and \,\ref{fig:colocolo_hoii}, we adopt a magnitude cut of $M_V =-3$. 
Therefore, we list no number and fraction of OBe stars for these dimmer sources in Table\,\ref{tab:nbe}.
As for Sextans\,A, for Ho\,I and Ho\,II the color cuts are again placed slightly conservatively, at the point where $N/N_\mathrm{max}$ reaches down to the value 1/7 in the number distributions shown above the color-color diagrams.

\subsection{Completeness \label{sec:appb_completeness}}
\paragraph{Magellanic Clouds} We estimate the completeness of the cross-matched data products by counting the number of sources they contain in different magnitude intervals, and dividing that number by the number of sources in the GAIA EDR3 catalog (after cleaning the GAIA EDR3 catalog of foreground stars in the same fashion).
The GAIA EDR3 catalog should be essentially complete \citep{Arenou18} in our magnitude range of interest ($ 12 \lesssim m_{Gbp} \lesssim 17$).
We only consider a part of the sky that is fully covered by the cross-matched data product (e.g, the $UB$ data do no cover the SMC's Wing). 

\begin{figure*}[ht]
\centering
\includegraphics[width=0.5\linewidth]{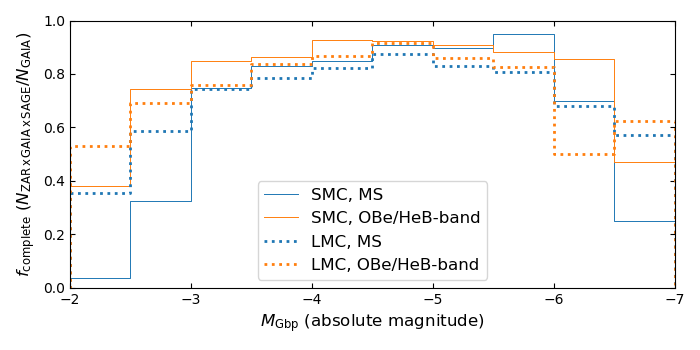}
\caption{Completeness fraction estimate for main sequence (MS) sources and sources in the band that contains OBe stars and He-burning sources (OBe/HeB-band) -- see text -- shown for both the Small Magellanic Cloud (SMC) and the Large Magellanic Cloud (LMC). }
\label{fig:f_complete}
\end{figure*}

With GAIA photometry alone we cannot tell apart MS, OBe, and HeB stars. Therefore, we estimate the completeness in the `MS band' and the `OBe/HeB band'. In the SMC, we define the MS band as the sources with $Gbp-Grp < -0.15$ and for the OBe/HeB band we adopt $-0.15 < Gbp-Grp < 0.2$. For the LMC, we place the cuts at $Gbp-Grp < -0.1$ for the MS band and $-0.1 < Gbp-Grp < 0.25$ for the OBe/HeB band.
The result is shown in Fig.\,\ref{fig:f_complete}.
We find that the completeness is about $80-90$\% in the magnitude range $-3 > M_\mathrm{Gbp} > -5.5$ (i.e., the range in which we plot the OBe star fraction in Fig.\,\ref{fig:fbe_both}). For the dimmest sources, the completeness drops dramatically, especially for sources in the MS bands. We interpret this as the consequence of the hot MS sources becoming too dim to be observed in the $[3.6]$ band. This drop is deeper for the SMC because its distance modulus is half a magnitude larger \citep{Macri06, Hilditch05}.

\begin{figure*}[ht]
\centering
\includegraphics[width=0.5\linewidth]{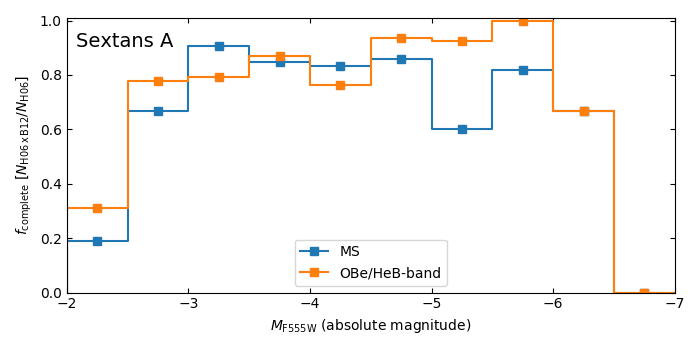}
\caption{Completeness fraction estimate for main sequence (MS) sources and sources in the band that contains OBe and He-burning stars (OBe/HeB-band) -- see text -- shown for Sextans\,A.}
\label{fig:f_complete_sexa}
\end{figure*}

\paragraph{Sextans A}
Similar to the SMC and LMC photometric data sets, we estimate the completeness of the cross-matched Sextans\,A HST data. Here, we adopt $F555W - F814W < -0.05$ for the MS band and $ -0.05 < F555W - F814W < 0.2$ for the OBe/HeB band. Now, we divide the number of sources in the cross-correlated H06 x B12 catalog by the number of sources with a $F555W$ as well as $F814W$ magnitude in H06. As such, this completeness estimate assumes that the H06 $F555W$ data are complete, which might not be the case. However, we see no clear reason for why the H06 $F555W$ data (that go two magnitudes deeper than the the dimmest sources we consider at $M_{F555W} = -2$, the limit in our CMD) would be significantly more (or less) complete for MS stars than for OBe stars, which would bias the values we derive for the OBe star fraction. Figure \ref{fig:f_complete_sexa} shows that the cross-correlated data set is about 80\% complete, compared to the H06 catalog. In the absolute magnitude range $-2 > M_{F555W} > -3$ (where we do not present an OBe star fraction) fewer sources are retained after the cross-correlation. 

\begin{figure*}[ht]
\centering
\includegraphics[width=\linewidth]{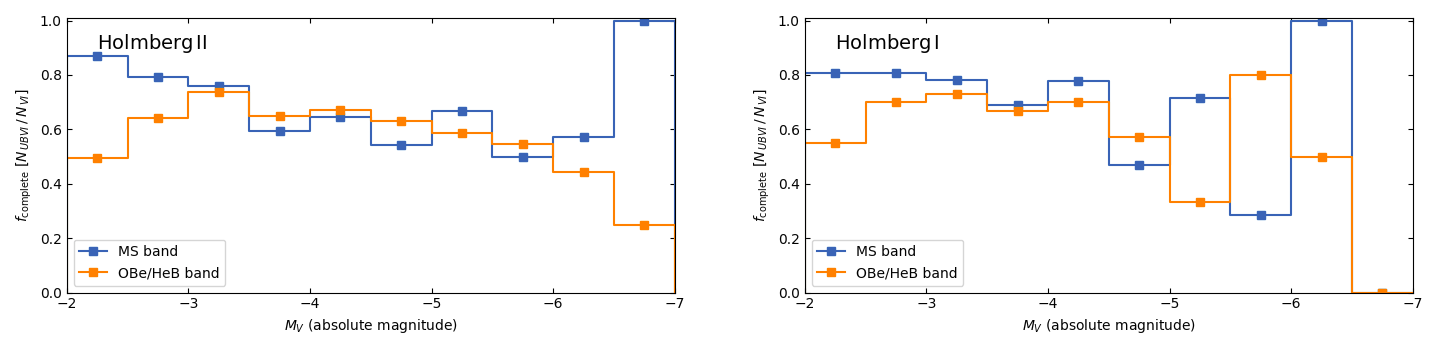}
\caption{Completeness fraction estimate for main sequence (MS) sources and sources in the band that contains OBe and He-burning stars (OBe/HeB-band) -- see text -- shown for Holmberg\,II (\textit{left}) and Holmberg\,I (\textit{right}).}
\label{fig:f_complete_hos}
\end{figure*}

\paragraph{Holmberg I and Holmberg II}
For Holmberg\,I and Holmberg\,II we adopt a similar approach as described above for Sextans\,A.
To estimate the completeness, here we divide the number of sources that have $UBVI$ data (i.e., in all four filters) by the number of sources that have at least $VI$ data, in different $M_V$ intervals. We adopt $V - I < -0.1$ for the MS band and $ -0.1 < V - I < 0.25$ for the OBe/HeB band in both Holmberg\,I an II. The resulting magnitude-dependent completeness fractions are shown in Fig.\,\ref{fig:f_complete_hos}. We find that in the range $-3 > M_V -5.5$, where we present OBe star fractions, the estimated completeness is about 70\%. Most importantly,  we obtain very similar values for the completeness in the MS band and the OBe/HeB band, implying that completeness issues do not bias our derived values for the OBe star fraction.

\subsection{Pollution by Wolf-Rayet and B[e] stars \label{sec:appb_pollution}}

\begin{figure*}[ht]
\centering
\includegraphics[width=0.65\linewidth]{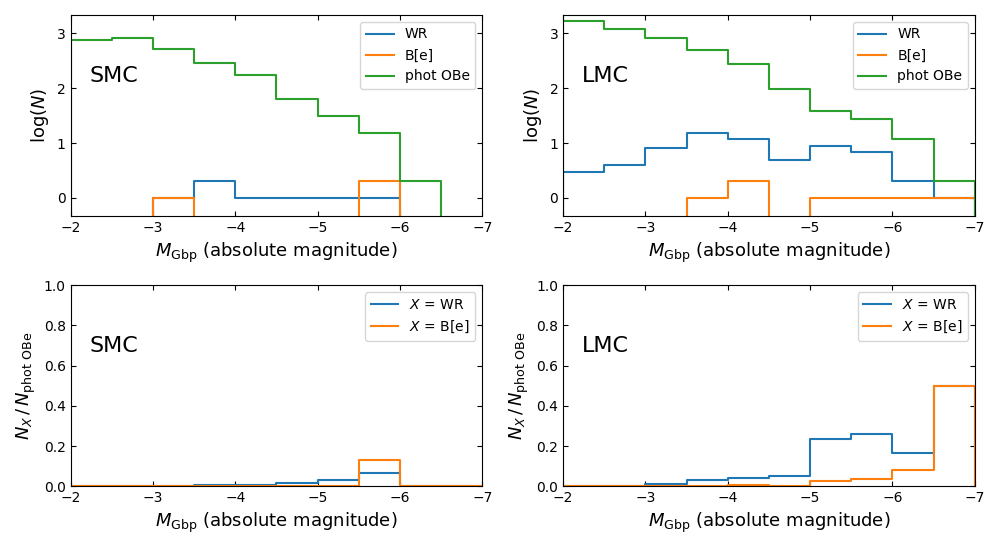}
\caption{\textit{Top panels:} number distributions of absolute magnitude $M_{Gbp}$ for various types of sources. The green line shows photometric OBe stars (`phot OBe'; as identified by the color cuts in Figs.\,\ref{fig:colocolo_3cut} and \ref{fig:colocolo_3cut_lmc}). The orange and blue lines show photometric OBe stars that are known to instead be B[e] and Wolf-Rayet (WR) stars, respectively, and that are in the SMC and LMC spectral type catalogs of \cite{Bonanos09, Bonanos10}.
\textit{Bottom panels:} fraction of all photometric OBe stars that is known to be a WR star or B[e] star instead, in different absolute magnitude intervals.}
\label{fig:wr_bbe_konte}
\end{figure*}

We use the B09 and B10 catalogs to investigate potential contamination of our OBe stars by Wolf-Rayet (WR) and B[e] stars in the Magellanic Clouds.
The LMC B09 catalog contains 89 WR stars in the magnitude range $-2 > M_{Gbp} > -7$. Out of these, 69 (i.e., the vast majority) have colors that would label them as OBe stars (see Fig.\,\ref{fig:colocolo_3cut_lmc}). The B[e] stars in the magnitude range $-2 > M_{Gbp} > -7$ are fewer in number (eight), and seven of these fall within the OBe color cut (typically they are far redder in the infrared than OBe stars, with $J-[3.6] > 2$). In the top right panel of Fig.\,\ref{fig:wr_bbe_konte}, we show the logarithm of the number of photometric OBe stars and the polluting WR and B[e] stars in the LMC. In the bottom right panel, we display the fraction of photometric OBe sources in the LMC that is known to have a WR or B[e] spectral type. It shows that for sources brighter than $M_{Gbp} = -5$, where the photometric OBe stars become few in number, polluting WR and B[e] stars start to make up a significant fraction. At least 20\% of the photometric OBe stars in the LMC brighter than $M_{Gbp} = -5$ are known to not be OBe stars. For dimmer sources, however, pollution by WR and B[e] stars plays no significant role. We also note that the current census of WR stars in the LMC and SMC should be close to complete \citep{Neugent18}. Therefore, we do not expect the WR pollution of photometric OBe stars to be significantly stronger than what is shown in the bottom panels of Fig.\,\ref{fig:wr_bbe_konte}.

In the SMC (left panels of Fig.\,\ref{fig:wr_bbe_konte}), there are relatively fewer WR stars than in the LMC. One of the reasons for this is that due to the weaker winds, stars need to be more luminous to manifest themselves as WR stars \citep{Shenar20}.
There are also three B[e] stars in the B10 catalog that have OBe-like colors in our magnitude range of interest. For the SMC, only in the $-6 < M_{Gbp} < -5.5$ bin there is a significant pollution by WR and B[e] stars (20\%).

From the above, we conclude that WR star sources more often than not manifest themselves as photometric OBe stars, but that they are far fewer in number than genuine OBe stars. The exception is that for sources brighter than $M_{Gbp} = -5.5$, the WR plus B[e] star pollution can be at least 20\% in both the SMC and the LMC. For this reason, we refrain from presenting an OBe star fraction for sources brighter than $M_\mathrm{abs} = -5.5$.

\subsection{Assessing uncertainties in calculating the OBe star fraction \label{sec:appb_uncertainties}}
To calculate the OBe star fraction in various absolute magnitude intervals, we divide the number of OBe stars ($N_\mathrm{OBe}$) by the total number of H-burning stars, which is the sum of the number of OBe stars plus the number of non-OBe MS stars ($N_\mathrm{MS}$): $f_\mathrm{OBe} = N_\mathrm{OBe} / (N_\mathrm{OBe} + N_\mathrm{MS}$).
Then, we calculate the statistical error as: $\sigma_{f_\mathrm{OBe}} = \sqrt{ f_\mathrm{OBe} \cdot \frac{ 1-f_\mathrm{OBe} }{N_\mathrm{OBe} + N_\mathrm{MS}}}$. This is a lower limit on the true error, as other uncertainties (e.g., the exact location of the MS-OBe border in our color-color diagrams) are not quantifiable, at least not in all our target dwarf galaxies. 

\begin{figure*}[ht]
\centering
\includegraphics[width=0.65\linewidth]{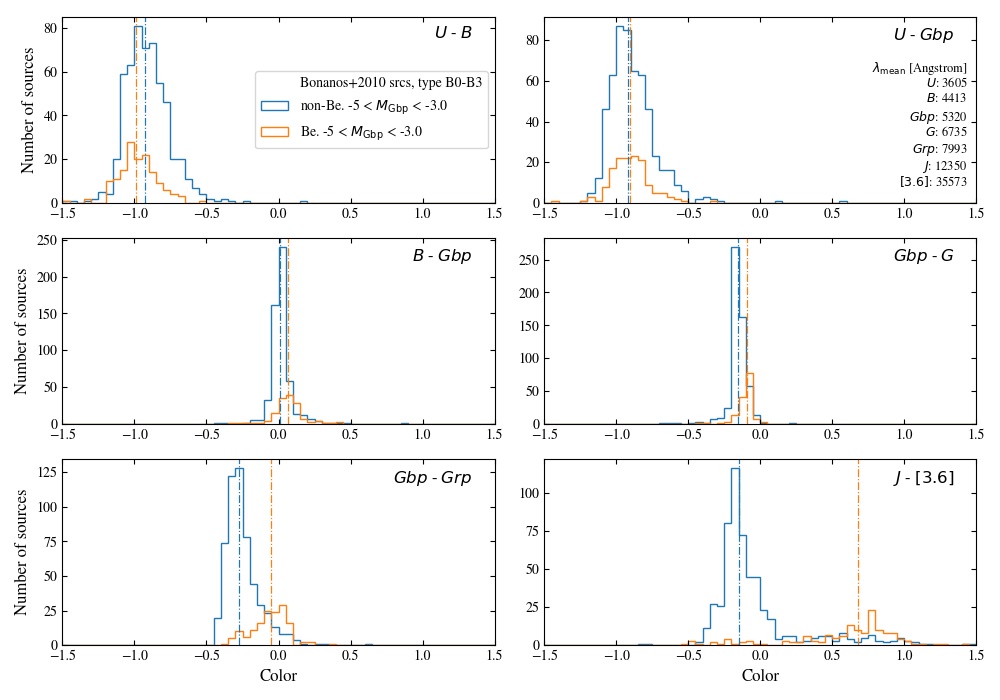}
\caption{Histograms of the different colors of sources that have a B spectral type \citep{Bonanos10} (including OBe). Only early B stars (B0 to B3) with an absolute $Gbp$ magnitude of $-3 > M_\mathrm{Gbp} > -5$ are included. The dash-dotted lines show the median color. The mean wavelength $\lambda_\mathrm{mean}$ of each filter (top right panel) is taken from \url{http://svo2.cab.inta-csic.es/theory/fps/}.}
\label{fig:Bcolors}
\end{figure*}

In theory, the disk of an OBe star could contribute to the $Gbp$, $V$, or $F555W$ flux, biasing our comparison. 
We do not correct for a disk contribution; below, we justify this. In Fig.\,\ref{fig:Bcolors}, we show the distribution of various colors for early B and Be sources in the B10 catalog. This figure shows that these B and Be stars look very similar in colors that are constructed with the bluer filters ($U-B$, $U-Gbp$, $B-Gbp$). We interpret this as being a result of an at most small disk contribution in the $U$, $B$, and $Gbp$ filters. This makes a correction for the disk contribution to $Gbp$ unnecessary. It is only in the $Gbp - Grp$ and infrared colors that OBe stars start to look relatively red. We note that to measure the OBe star fractions we chose relatively blue filters (i.e., $Gbp$, $V$, and $F555W$ instead of $G$/$Grp$, $I$, and $F814W$) on purpose, because there we expected the disk contribution to play a less significant role.

An uncertainty for the OBe fraction could be an apparent lack of young stars that are observable. In the SMC, it was found that massive stars in the first half of their lifetime seem to be making up far less than half of the population \citep{Schootemeijer21}, possibly because they are hidden by their birth cloud, or perhaps as a result of a declining star-formation rate. Stars at low $Z$ in the second half of their lifetime are predicted to be OBe stars more often -- as a result of core contraction in the single star channel \citep{Ekstrom08, Hastings20}, and binary interaction typically not taking place during the early MS phase \citep{Wang20}. Therefore, a dearth of stars in the first half of their MS lifetime could cause us to overestimate the OBe star fraction in the SMC, at least compared to a scenario where star formation is constant and stars are observable all their life. However, we do not expect the star formation histories of the other dwarf galaxies to follow the one of the SMC, indicating that the high OBe star fraction of $\sim$0.3 is not an artefact caused by star formation history.

\section{Magnitude-mass relations for blue stars \label{sec:appc}}
\begin{figure*}[ht]
\centering
\includegraphics[width=0.75\linewidth]{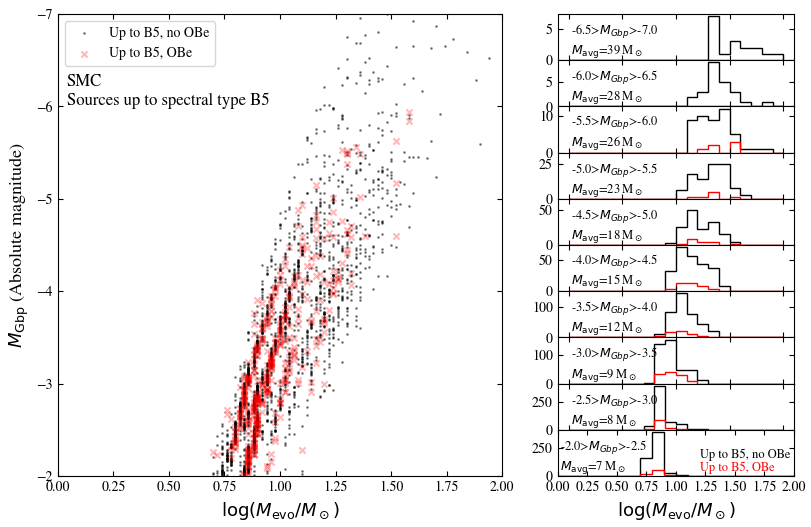}
\caption{\textit{Left:} diagram showing the evolutionary mass $M_\mathrm{evo}$ of stars with spectral types B5 and earlier plotted against absolute magnitude $M_{Gbp}$, for both stars with (OBe) and without (MS) emission features in their spectra. The spectral type catalog is from \cite{Bonanos10} and the evolutionary masses were determined by \cite{Schootemeijer21}. \textit{Right:} histograms that show the evolutionary mass distributions of the sources that are displayed in the left panel, in different $M_{Gbp}$ intervals. The average evolutionary masses of sources in these magnitude intervals (i.e., OBe and MS) are written in each panel.}
\label{fig:mevo}
\end{figure*}

To estimate stellar masses for OBe and non-OBe MS stars from their absolute magnitudes, we look up the evolutionary masses that have been determined for these sources in the SMC by \cite{Schootemeijer21}, to which we refer below as S21. From S21, we consider only stars that have spectral types B5 and earlier, and we include both non-OBe and OBe stars.
In the left panel of Fig.\,\ref{fig:mevo} we plot their evolutionary mass against absolute magnitude, and on the right we show the evolutionary mass distribution in different absolute magnitude intervals.
We note that there might be other biases resulting from selection effects in the B10 catalog.
However, the B10 catalog is expected to be rather complete -- about 80\% at the bright end, and roughly 50\% for dimmer sources, according to S21.
Another possible source of error in the magnitude-mass relations comes from extrapolating them from the SMC to other galaxies and other filters ($F555W$ and $V$ instead of $Gbp$). We note that a difference of $\Delta V = 0.5$\,mag has been inferred between equal-mass massive stars in the Milky Way and at Z$_\odot / 50$ \citep{Evans19}, implying offsets of the order of a tenth of a magnitude in the metallicity range of $0.1 \leq Z / Z_\odot < 0.5$ considered in this work.
Because of this, in combination with the sizeable scatter of evolutionary masses in each $M_{Gbp}$ interval, we caution that the interpretation of our $M_{abs}$ - evolutionary mass relations for blue stars should not be that each of the MS and OBe sources has exactly the mass given by the relation; instead the masses shown at the top of Fig.\,\ref{fig:fbe_both} are intended to serve as an indication.

\section{Simulations: comparison with theory in the Small Magellanic Cloud}
\subsection{Single stars \label{sec:app_ss} }
We first theoretically predict the OBe star fraction in the SMC under the assumption that all OBe stars are formed through the single star channel, that is, without the aid of a binary companion. For this, we repeat the star cluster study of \cite{Hastings20}, but now assuming constant star formation instead of a coeval scenario and extending the mass range of the models to 60\,M$_\odot$. This upper mass limit sufficiently high because more massive stellar models are brighter than $M_{Gbp} = -5$ their entire lifetime. This analysis adopts an initial rotation distribution based on observations \citep{Dufton13}.
We note that \cite{Dufton13} assume random spin axis orientation. Alignment of spin axes \cite[as found by][in old open clusters]{Corsaro17} could affect the distribution of \cite{Dufton13}, although other work does support random spin axis orientation \citep{Hummel99, Jackson10, Mosser18, Gehan21}.

To obtain the absolute $Gbp$ magnitude, we use a model star's effective temperature and luminosity, and bolometric corrections from MIST \citep[\url{http://waps.cfa.harvard.edu/MIST/model_grids.html} --][]{Dotter16, Choi16}. The result is shown in Fig.\,\ref{fig:fbe_smc_vs_theory}.

\subsection{Binary stars \label{sec:app_bin}}
We also theoretically predict the OBe star fraction under the assumption that only binary systems produce OBe stars. We simulate a population of binary systems with the rapid binary evolution code ComBinE \citep{Kruckow18}, and briefly highlight the most important physics assumptions here. We again adopt a constant star formation rate. The initial rotation velocities are chosen such that the rotation periods of both stars equals the orbital period, as is the case for tidally synchronized systems. We adopt a Salpeter Initial Mass Function and an initial mass range of $5 < M_\mathrm{ini}/M_\odot < 60$.
The initial mass ratio distribution is flat and it ranges from $q_\mathrm{ini}=0.1$ to $q_\mathrm{ini}=1$. In the orbital period distribution the number of systems scales with the logarithm of the orbital period to the power $\pi = -0.55$ \citep{Sana12}. The initial orbital period range is $0.15 < \log (P_\mathrm{orb,\, ini}/d) < 3.5$. For the extreme initial mass ratios of $q < 0.3$, we assume that the stars merge as Roche lobe overflow (RLOF) commences. We also assume that RLOF in systems where the donor has a convective envelope (that goes deeper than 10\% of the stellar mass) leads to a merger, and that contact systems merge.
This merger product does remain in the stellar population. For the merger product we assume that 10\% of the combined stellar mass is lost in the merger event. 
The starting point of the merger product is a single star model of the appropriate mass (i.e., 90\% of the combined progenitor star masses) that has burned the same amount of hydrogen as both progenitors stars have burned pre-merger.
In our simulations the merger does not produce an OBe star because it is expected that a high amount of angular momentum is lost post-merger \citep{Schneider19}.
All systems that do not meet one of the merger criteria described above are assumed to go through stable RLOF, resulting in the accretor becoming an OBe star. We assume that the material that is lost from the system during RLOF carries the same specific orbital angular momentum as the accretor. Tidally induced changes of the stellar spins are accounted for. 
We consider these assumptions to be optimistic, that is, favoring a high OBe star fraction from the binary channel. 

\begin{figure*}[ht]
\centering
\includegraphics[width=0.45\linewidth]{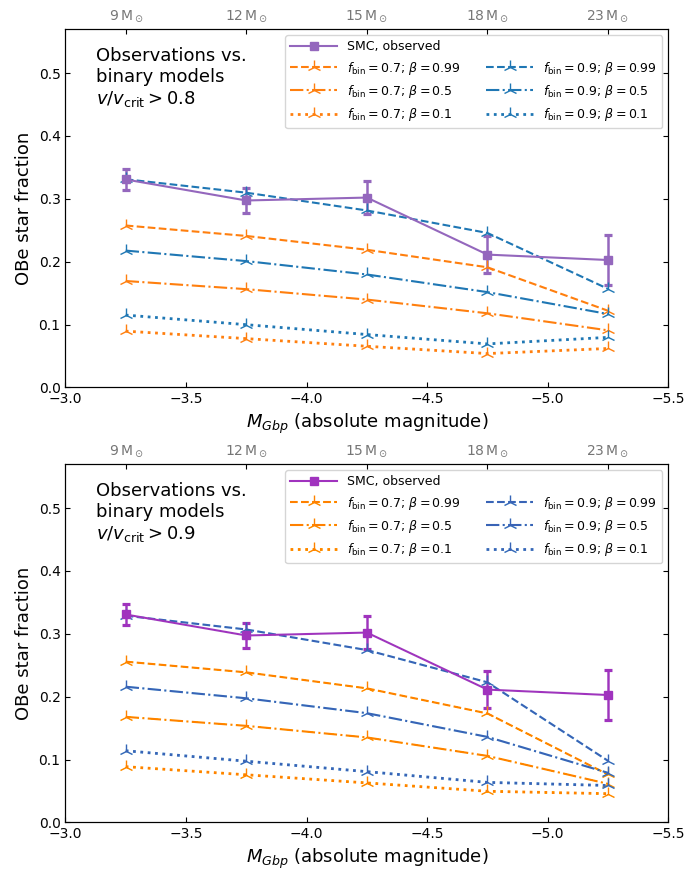}
\caption{
Same as the lower panel of Fig.\,\ref{fig:fbe_smc_vs_theory}, but here $v/v_\mathrm{crit} > 0.8$ and $v/v_\mathrm{crit} > 0.9$ are set as the required rotation velocity stars to be OBe stars in the binary populations .}
\label{fig:fbe_smc_vs_theory_vvcbin0809}
\end{figure*}

When we adopt a binary fraction lower than $f_\mathrm{bin} = 1$, we assume that the single stars never become OBe stars, and simply calculate the OBe star fraction as $f_\mathrm{OBe} = f_\mathrm{bin} \cdot f_\mathrm{OBe\,(f_\mathrm{bin}=1)}$, the latter term being the OBe star fraction calculated for a binary fraction of unity. For binaries, we show the magnitude-dependent theoretical OBe star fractions at the bottom half of Fig.\,\ref{fig:fbe_smc_vs_theory}. 
These are calculated under the assumption that stars rotating at $v/v_\mathrm{crit} \gtrsim 0.7$ manifest themselves as OBe stars. Figure\,\ref{fig:fbe_smc_vs_theory_vvcbin0809} shows the OBe star fractions that we obtain when $v/v_\mathrm{crit} \gtrsim 0.8$ and 0.9 are set as minimum rotation rates. A higher minimum rotation rate for OBe stars slightly reduces the number of OBe stars in the theoretical binary population, but not nearly as much as for the single stars.

\end{document}